\documentclass[twocolumn,showpacs,epsfig]{revtex4}
\usepackage{epsfig,amsmath}

\newcommand{\kap}[1][ ]{K$^{+}${#1}}
\newcommand{\kam}[1][ ]{K$^{-}${#1}}

\newcommand{\AGeV}[1][ ]{$A$~GeV{#1}}
\newcommand{\Apart}[1][ ]{$A_{\rm part}${#1}}
\begin{document}

\title{Production of $\bf{K^{\boldsymbol{+}}}$ and of
$\bf{K^{\boldsymbol{-}}}$
Mesons in Heavy-Ion Collisions\\
from 0.6 to 2.0~$\bf{A}$~GeV Incident Energy}

\author{
A.~F\"orster$^{b,*}$,
F.~Uhlig$^{b,a}$, 
I.~B\"ottcher$^d$, 
D.~Brill$^a$,
M.~D\c{e}bowski$^{e,f}$, 
F.~Dohrmann$^f$, 
E.~Grosse$^{f,g}$,
P.~Koczo\'n$^a$, 
B.~Kohlmeyer$^d$, 
S.~Lang$^{b,a}$,
F.~Laue$^{a,+}$, 
M.~Mang$^a$, 
M.~Menzel$^d$, 
C.~M\"untz$^{b,a,c}$,
L.~Naumann$^f$, 
H.~Oeschler$^b$, 
M.~P{\l}osko\'n$^{a,c}$,
W.~Scheinast$^f$, 
A.~Schmah$^{b,a}$, 
T.~J.~Schuck$^{c,*}$,
E.~Schwab$^a$, 
P.~Senger$^a$, 
Y.~Shin$^c$, 
J.~Speer$^{a,\dag}$,
H.~Str\"obele$^c$, 
C.~Sturm$^{b,a,c}$, 
G.~Sur\'owka$^{a,e}$,
A.~Wagner$^{f,b}$,
W.~Walu\'s$^e$\\
(KaoS Collaboration)\\
$^a$ Gesellschaft f\"ur Schwerionenforschung, D-64220 Darmstadt, Germany\\
$^b$ Technische Universit\"at Darmstadt, D-64289 Darmstadt, Germany\\
$^c$ Johann Wolfgang Goethe-Universit\"at, D-60325 Frankfurt am Main, Germany\\
$^d$ Phillips-Universit\"at, D-35037  Marburg, Germany\\
$^e$ Uniwersytet Jagiello{\'n}ski, PL-30059 Krak\'ow, Poland\\
$^f$ Forschungszentrum Dresden-Rossendorf, D-01314 Dresden, Germany \\
$^g$ Technische Universit\"at Dresden, D-01062 Dresden, Germany\\
$^*$ Present address: Max-Planck-Institut f\"ur Kernphysik,
     D-69117 Heidelberg,  Germany \\
$^+$ Present address: Brookhaven National Laboratory, Upton, NY 11973, USA \\
$^\dag$ deceased
}

\begin{abstract}

This paper summarizes the yields and the emission patterns
of \kap and of \kam mesons measured in inclusive \mbox{C+C},
\mbox{Ni+Ni} and \mbox{Au+Au} collisions at incident energies from
$0.6$~\AGeV to $2.0$~\AGeV using the Kaon Spectrometer KaoS at GSI.
For \mbox{Ni+Ni} collisions at $1.5$ and at $1.93$~\AGeV as well
as for \mbox{Au+Au} at $1.5$~\AGeV detailed results of the
multiplicities, of the inverse slope parameters of the 
energy distributions and of the
anisotropies in the angular emission patterns as a function of the
collision centrality are presented. 
%By comparing the measured \kap production yields
%data to transport-model calculations a compression modulus of
%$K_{\rm N} \approx 200$~MeV for the nuclear equation of state is
%extracted. 
 When comparing transport-model calculations to the measured 
\kap production yields an agreement is only obtained 
for a soft nuclear equation of state (compression modulus
$K_{\rm N} \approx 200$~MeV). 
The production of \kam mesons at energies around $1$ to
$2$~\AGeV is dominated by the strangeness-exchange reaction
$\rm{K}^- \rm{N} \rightleftharpoons \pi \rm{Y}$ ($\rm{Y}=\Lambda,
\Sigma$) which leads to a coupling between the \kam and the \kap
yields. However, both particle species show distinct differences
in their emission patterns suggesting different freeze-out
conditions for \kap and for \kam mesons.

\end{abstract}


version of \today

\maketitle

\section{Introduction}
\label{introduction}

Relativistic heavy-ion collisions at incident energies ranging
from $0.6$ to $2.0$~\AGeV provide a unique opportunity to study
the behavior of nuclear matter at high densities. These studies
are important challenges for testing the present understanding of
nuclear matter. In addition, they are of relevance for astrophysics,
as the modelling of neutron stars or of supernovae depends on the
properties of nuclear matter under these extreme conditions~\cite{SN}.

In central \mbox{Au+Au} collisions at the incident energies under
investigation
densities of 2 -- 3 times normal nuclear matter density can be reached
\cite{Hart_NP94,Fuchs_Rev,hart_habil}. A sensitive probe to test
these conditions is the production of strange mesons at or below
the production thresholds of these particles in free NN
collisions. The rest mass of charged kaons is 0.454 GeV.
For the \kap production the threshold in NN collisions is 1.58 GeV (in the
laboratory system) as defined by the associate production $\rm{NN}
\rightarrow \rm{K}^+ \Lambda \rm{N}$ and it is 2.5 GeV for the
\kam production via pair creation $\rm{NN} \rightarrow \rm{NN}
\rm{K}^- \rm{K}^+$.

The key mechanism for the \kap production in heavy-ion reactions
at these incident energies is the accumulation of the necessary
energy by multiple collisions of particles inside the reaction
zone. Higher densities increase the number of these collisions and
especially second generation collisions like $\Delta \rm{N} $ with
sufficiently high relative momentum to create a \kap occur most
frequently during the high-density phase of the reaction. The
density reached in the reaction zone depends on the stiffness of
nuclear matter. Because of their specific production mechanism and
because of their rather long mean free path ($\approx 5$~fm
at normal nuclear density) \kap mesons are
ideal probes to explore the high-density phase of a heavy-ion
reaction and to study the stiffness of the nuclear equation of
state (EoS)~\cite{ko84,aich,sturm,fuchs,Hart_eos}.

In contrast, the
behavior of \kam mesons in a dense nuclear medium is expected to
be very different from the one of the \kap mesons due to two distinct
properties:

(i) The interaction with nuclear matter: The \kap are hardly
absorbed in nuclear matter due to strangeness conservation. It is
very unlikely that a rare \kap (containing an $\bar s$ quark)
encounters an equally rare hyperon Y ($\Lambda$, $\Sigma$)
containing an $s$ quark. The \kam on the contrary, can easily be
absorbed on a nucleon converting it into a hyperon and a pion.
Consequently, the mean free path of the \kam is significantly
shorter than the one of the \kap[].
The strangeness-exchange reaction $\rm{K}^- \rm{N}
\rightleftharpoons \pi \rm{Y}$ has a large cross section and is
therefore responsible for the appearance and disappearance of
\kam mesons. This channel has been suggested to be the dominant
production mechanism in nucleus-nucleus collisions~\cite{ko84} and
this has been demonstrated in \cite{AF,Oe00,Hart03}.

(ii) The influence of KN potentials:
According to various theoretical approaches
the KN interaction is governed by the superposition
of a scalar and a vector potential \cite{kaplan,gbrown,waas,schaf,lutz,laura}.
While the scalar potential acts attractively on both kaon species, the vector
potential repels \kap and attracts \kam[]. For \kap these two
contributions almost cancel leading to a small repulsive $\rm{K}^+\rm{N}$
interaction. For the \kam the addition of both attractive
interactions results in a strongly attractive potential.
Attempts to observe these effects in the respective production
cross sections are under
discussion~\cite{laue,menzel,crochet,wisniewski,brown,cass_brat,Hart03}.
These potentials are predicted to have a
sizeable effect on the azimuthal emission patterns of \kap and of
\kam (elliptic flow) \cite{li_v2,wang_v2} which has been observed at the
Kaon Spectrometer \cite{shin,FU}.

This paper intends to give a comprehensive overview
on the cross sections and
on the emission patterns of the \kap and of the \kam production in
mass-symmetric heavy-ion reactions in the incident energy range from
$0.6$ to $2$~\AGeV[].
It summarizes new as well as previously
published~\cite{laue,laue_epj,menzel,sturm,AF} results on
inclusive collisions of  \mbox{Au+Au}, \mbox{Ni+Ni} and \mbox{C+C}
collisions.
Furthermore, new results focussing on the centrality
dependence of \mbox{Ni+Ni} collisions at $1.5$ and at
$1.93$~\AGeV and of \mbox{Au+Au} collisions at $1.5$~\AGeV
are presented.
Results on the kaon production in the
mass-asymmetric collision systems \mbox{C+Au} and \mbox{Au+C} have
been published recently \cite{schmah}. The results on the
azimuthal distributions of kaons~\cite{shin,FU} as well as of
pions and of protons~\cite{Brill} have been published and further
publications on this topic are in preparation
as well as a review on pion production.

This paper is structured in the following way: First, in
Section~\ref{experiment}, a description of the experimental setup
and of the data analysis is given. In
Section~\ref{results_inclusive} we summarize the results on cross
sections, energy distributions, and polar angle distributions for
inclusive collisions, i.e.~without any selection in the collision
centrality. Section~\ref{results_centrality} presents a detailed
study of the centrality dependence of the \kap and of the \kam
production. In Section~\ref{discussion_coupling_kmkp} the measured
yields of \kap and of \kam mesons are discussed showing that their
production is correlated. Despite this correlation of the
production yields the emission patterns of \kap and of \kam mesons
differ significantly. Section~\ref{discussion_freezeout} discusses
these differences with respect to the influences of the
KN-potentials and with respect to different emission times of the
two kaon species. Section~\ref{discussion_eos} compares the
production yields of \kap mesons in different collision systems to
recent transport model calculations to extract information on the
stiffness of the nuclear equation of state.

\section{The Experiment}
\label{experiment}

\subsection{The Setup}
\label{experiment_setup}

The experiments were performed with the Kaon Spectrometer (KaoS)
at the heavy-ion accelerator SIS (Schwerionensynchrotron)  at the
GSI (Gesellschaft f{\"u}r
Schwerionenforschung) in Darmstadt, Germany. A detailed
description is presented in \cite{senger}. Here we just briefly review
the main features.

\begin{figure}
\epsfig{file=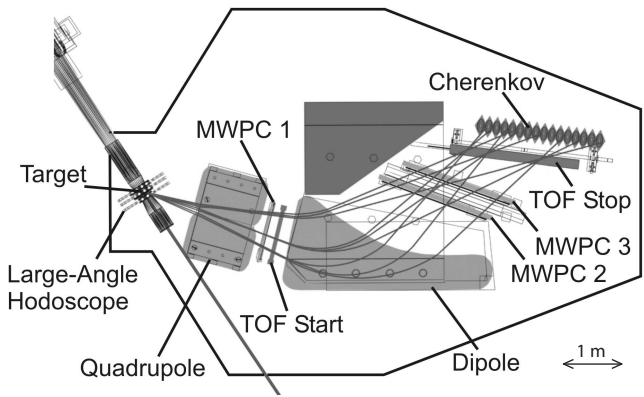,width=8.5cm} \caption{Top view of the
Kaon Spectrometer KaoS with its various detector components.}
\label{KAOS3}
\end{figure}

The setup of the quadrupole-dipole spectrometer KaoS is shown in
Fig.~\ref{KAOS3}. Positively and negatively charged particles are
measured separately using different magnetic field polarities. The
magnetic spectrometer has an acceptance in solid angle of $\Omega
\approx 30$~msr and covers a momentum bite of $p_{\rm max}/p_{\rm
min} \approx 2$. The short distance of $5$ - $6.5$~m from the
target to the focal plane minimizes the number of kaon decays in
flight. The loss of kaons by decay and due to the geometrical
acceptance is accounted for by corrections which are determined by
Monte-Carlo simulations using the code GEANT~\cite{GEANT}.
Particle identification is based on momentum and on time-of-flight
(TOF) measurements. The trigger system is as well based on the
time-of-flight information to separate pions, kaons and protons.
For the separation of high momentum protons from kaons a threshold
Cherenkov detector \cite{misko_nim} is used in addition. The
trigger system suppresses pions and protons by factors of 10$^2$
and of 10$^3$, respectively.

In total there are three time measurements using segmented plastic
scintillator arrays: The TOF Start detector between the quadrupole
and the dipole (16 modules), the TOF Stop detector in the focal
plane of the spectrometer (30 modules), and the Large-Angle
Hodoscope (LAH) around the target point covering polar laboratory
angles of $12^{\circ} \le \theta_{\rm{lab}} \le 48^{\circ}$ (84
modules). The latter allows for a second time-of-flight
measurement for background rejection and for the determination of
the collision centrality using the number of measured charged
particles.

The trajectory reconstruction is based on three large-area
Multi-Wire Proportional Counters (MWPC 1 - 3)~\cite{stelzer}, one
between the quadrupole and the dipole and two behind the dipole
magnet, each of them measuring two spatial coordinates.  The
efficiencies for kaon detection are larger than 95\% for each of
these detectors.

The beam intensity is monitored using two scintillator telescopes
positioned at backward angles ($\theta_{\rm{lab}} = \pm
110^{\circ}$), measuring the flux of charged particles produced in
the target which is proportional to the beam intensity. The
absolute normalization is obtained in separate measurements at low
beam intensities using a plastic scintillation detector directly
in the beam line. The beam intensities have been chosen such, that
the efficiency of the data acquisition system (DAQ) due to dead
time was always above 50\%.

\begin{figure}
\epsfig{file=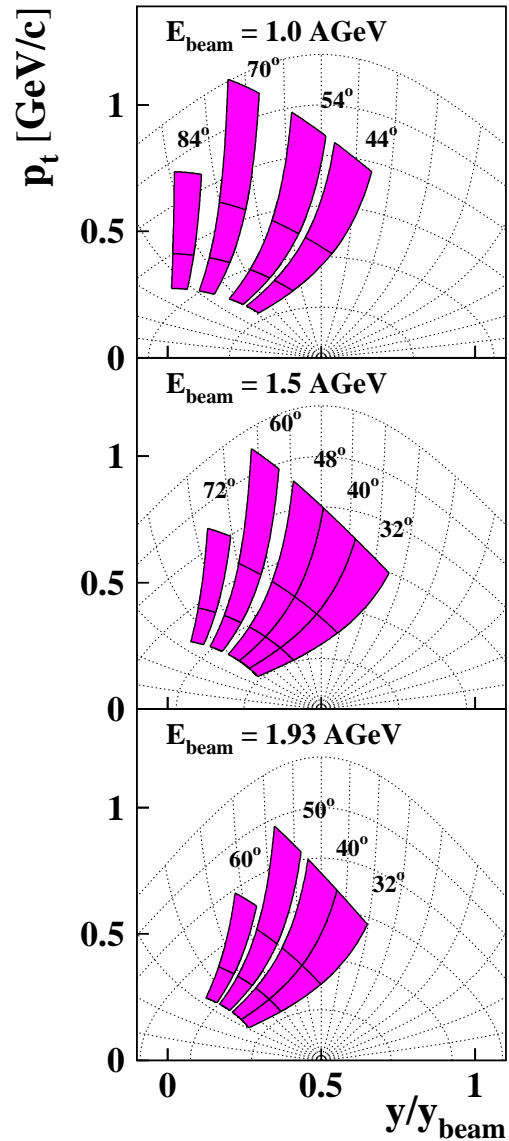,width=8cm}
 \caption{(Color online) Examples for the coverage of the Kaon Spectrometer
  in transverse momentum
  $p_{\rm t}$ and in normalized rapidity $y/y_{\rm beam}$ 
  for several laboratory angles
  $\theta_{\rm{lab}}$ (as indicated) and for various magnetic
  field settings at $1.0$~\AGeV[], at $1.5$~\AGeV[],
  and at $1.93$~\AGeV incident energy.}
\label{pty_kap}
\end{figure}

The spectrometer is mounted on a platform which can be rotated
around the target point on an air cushion in a polar angel range
from $\theta_{\rm{lab}} = 0^{\circ}$ to $130^{\circ}$. The angular
range covered at each position is $\Delta \theta_{\rm{lab}} = \pm
4^{\circ}$. Throughout this paper we will always quote the mean
value. The momentum coverage is maximized by measuring different
magnetic field settings ($|B_{\rm dipole}| = 0.6$, $0.9$, and
$1.4$~T). The resulting coverage for kaons in rapidity normalized
to the beam rapidity $y/y_{\rm beam}$ and in transverse momentum
$p_{\rm t}$ is sketched in Fig.~\ref{pty_kap} for three different
beam energies ($1.0$, $1.5$, and $1.93$~\AGeV). The shaded areas
correspond to different angular settings $\theta_{\rm{lab}}$ of
the spectrometer in the laboratory as denoted in the figure and to
various magnetic field settings.

In this paper we report on measurements of the collision systems
\mbox{C+C} ($0.8$, $1.0$, $1.2$, $1.5$, $1.8$, and $2.0$~\AGeV
beam energy), \mbox{Ni+Ni} ($1.1$, $1.5$, and $1.93$~\AGeV[]), and
\mbox{Au+Au} ($0.6$, $0.8$, $1.0$, $1.135$, and $1.5$~\AGeV[]).
The targets as well as their respective thicknesses and
interaction probabilities are given in Table \ref{table1}.
\begin{table}
\begin{center}
\begin{tabular}{c c c}
\hline target  &  thickness [mm] & interact. prob.\\
\hline
C & 3.0 & 2.7\% \\
Ni & 0.8  & 2.1\% \\
Au ($0.6$,$1.0$,$1.135$~\AGeV[]) & 1.0 & 3.6\% \\
Au ($0.8$,$1.5$~\AGeV[]) & 0.5  & 1.8\% \\
\hline
\end{tabular}
\end{center}
\caption{Thicknesses and interaction probabilities for the targets used in
the various experiments.}
\label{table1}
\end{table}
Due to the energy loss in the Au targets the average effective
beam energies $E_{\rm beam}^{\rm eff}$ for kaon production in
these cases are $0.56$, $0.78$, $0.96$, $1.1$, and $1.48$~\AGeV[]. For the
other target materials the energy loss is negligible. Throughout the
text we use the values of the nominal beam energies $E_{\rm
beam}$. However, in all figures displaying data as a function
of the beam energy the data points are plotted at $E_{\rm beam}^{\rm eff}$.

In the case of the $1.5$~\AGeV Au beam an exceptional operation
of the GSI accelerator facility was required: acceleration of
$^{197}\rm{Au}^{63+}$ ions with the synchrotron SIS up to an
energy of $0.3$~\AGeV[], then extraction and full stripping,
followed by injection into the Experimental Storage Ring (ESR)
where the beam was cooled (electron cooling) and finally
re-injection into the SIS and acceleration up to $1.5$~\AGeV[].

\subsection{Data Analysis}
\label{experiment_analysis}

\subsubsection{Track Reconstruction}

For the reconstruction of particle tracks in the spectrometer,
reconstruction functions correlating the
spatial coordinates measured by different detectors are used to
combine the information of different detectors
to track candidates. For example, one
reconstruction function is used to calculate the $x$-coordinate in
the MWPC between the quadrupole and the dipole ($x_1$) as a
function of the $x$-coordinates of the two MWPCs behind the dipole
($x_2,x_3$). By comparing  a measured position in one detector
with the calculated position  based on the hits in other detectors
using the reconstruction functions,
track candidates are created. This is  done for the spatial
coordinates in the MWPCs as well as for the assignment of the
modules of the TOF detectors.

The reconstruction functions are determined
by Monte-Carlo simulations using a complete
description of the experimental setup within GEANT. Single tracks
are followed through the spectrometer and the correlation
functions are determined by fitting polynomial functions up to
seventh order and with up to three $x$-coordinates and up to three
$y$-coordinates to all these simulated tracks.

After the reconstruction a
resulting track candidate consists of $x$- and $y$-coordinates in
all the three MWPCs, the module numbers, the time and the
energy-loss information of the TOF Start and of the TOF Stop
detector as well as the time and the multiplicity information of
the Large Angle Hodoscope.

To determine the efficiency of the tracking procedure GEANT is used.
This time not only single tracks are generated but combinations
of one or several tracks and additional background hits in the different
detectors as observed in the experiment.
The resulting efficiencies
vary with the laboratory angle, the magnetic field strength, and
the size of the collision
system because of varying track and background multiplicities. 
The resulting efficiencies of the tracking procedure are 
always larger than 90\%.

For each track candidate constructed in the preceding steps, the
momentum $p_{\rm lab}$ and the length of the flight path
$\mit{\Delta} l$ between the TOF Start and the TOF Stop detector
modules are obtained from a lookup table generated with GEANT.
From these quantities together with the measured time difference
$\mit{\Delta} t$ between the two TOF detectors, the squared mass
over charge ratio $(m/Z)^2$ is calculated using
\begin{equation}
\left(\frac{m}{Z}\right)^2=\left(\frac{p_{\rm lab} \cdot c}{Z}\right)^2
\left[ \left(\frac{\mit{\Delta} t\cdot c}{\mit{\Delta} l}\right)^2 -1 \right].
\label{equation:tracking:4}
\end{equation}
Since the particles under investigation have $Z=1$ we simply use
'mass' in the following.

\subsubsection{Background Reduction and Cross Section Calculation}

\begin{figure}
 \epsfig{file=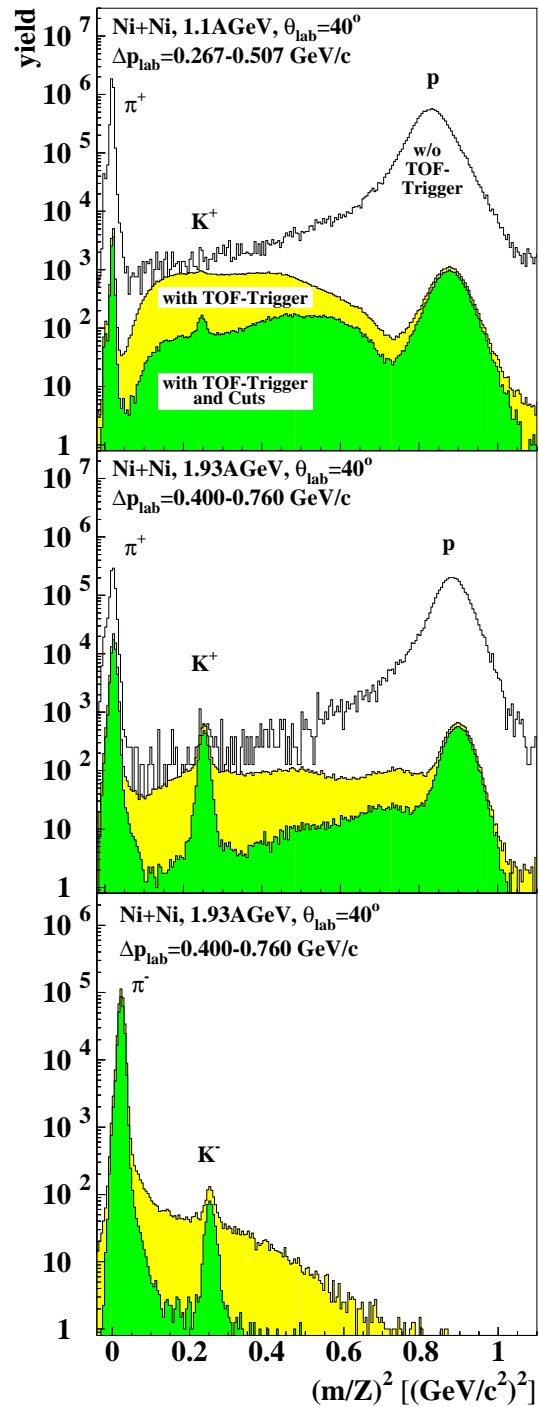,width=7.5cm}
 \caption{(Color online) Distributions of the
           squares of the masses assigned to the reconstructed tracks in
           three different cases: \mbox{Ni+Ni} collisions at $1.1$~\AGeV
           (upper panel), at $1.93$~\AGeV (middle panel),
           both using the field polarity for
           positively charged particles, and at $1.93$~\AGeV
           for negatively charged particles (lower panel). Differently shaded
           areas show the impact of the TOF trigger and of the application of
           the selection criteria during the data analysis. Details see
           text.}
 \label{mass_dis}
\end{figure}

Figure~\ref{mass_dis} shows mass distributions
for \mbox{Ni+Ni} collisions at $\theta_{\rm lab} = 40^{\circ}$.
The distribution in the upper panel has been measured at a beam
energy of $1.1$~\AGeV and at particle momenta $p_{\rm lab} = 0.267
\, - \, 0.507$~GeV/$c$ using the magnetic field setting for
positively charged particles. In this case of low beam energy and
of low particle momenta the region of the kaon mass is dominated by
background. The distribution labelled ``w/o TOF-Trigger'' was
measured using trigger conditions that forced every particle
passing through the spectrometer to be recorded.
The distribution labelled ``with TOF-Trigger'' shows the clear
reduction of pions and protons being recorded when using the 
time-of-flight trigger, but
still no clear \kap signal can be seen. The situation is different at
higher beam energies or at higher particle momenta where the kaons
are clearly visible already in the distributions
``with TOF-Trigger'' as can be seen
in the two lower panels of Fig.~\ref{mass_dis}. They show mass
distributions for \mbox{Ni+Ni} at $1.93$~\AGeV for positively
charged particles in the middle panel and for negatively charged
particles in the lower panel. In the latter case, the trigger condition
 ``w/o TOF-Trigger'' was not measured.

To reduce the remaining background effectively without loosing too
many kaons two types of selection criteria are applied:

(i) The so-called ``geometrical cuts'': These selection criteria
are based on the comparison between measured positions in one of the
MWPCs and
those extrapolated from hits in the other two using the
reconstruction functions described in the previous section. These
selection criteria are adjusted using measurements during the same
experiment which are nearly free of background (highest beam
energy and/or largest laboratory angle). This can be done since
the particle trajectories inside the spectrometer
only depend on the magnetic field and the geometrical
setup but not on quantities like $\theta_{\rm lab}$ or $E_{\rm beam}$.

(ii) The so-called ``velocity cut'': For each track candidate the
particle velocity between the TOF Start detector and the TOF Stop
detector is calculated as well as the velocity between the
Large-Angle Hodoscope and the TOF Stop detector. The comparison of
these two velocities is a powerful tool to suppress background
created by fake track candidates.
The ``velocity cut'' is adjusted using the measurements nearly free of
background as in the case of the ``geometrical cuts''.

Depending on the initial signal-to-background ratio the strength
of these cuts is varied between $5 \sigma$ and $2 \sigma$. After
the application of the selection criteria the final
signal-to-background ratio varies between $0.7$ and $120$. The
impact of the cuts on the mass distributions is as well shown in
Fig.~\ref{mass_dis} (labelled ``with TOF Trigger and Cuts''). In
many cases the remaining background is rather small as can be
clearly seen in the two lower mass distributions measured at
$1.93$~\AGeV[]. Even in the case of low beam energy and low
particle momenta a clear \kap signal can be observed after
applying the selection criteria (upper panel, please note the
logarithmic scale).

\begin{figure}
\epsfig{file=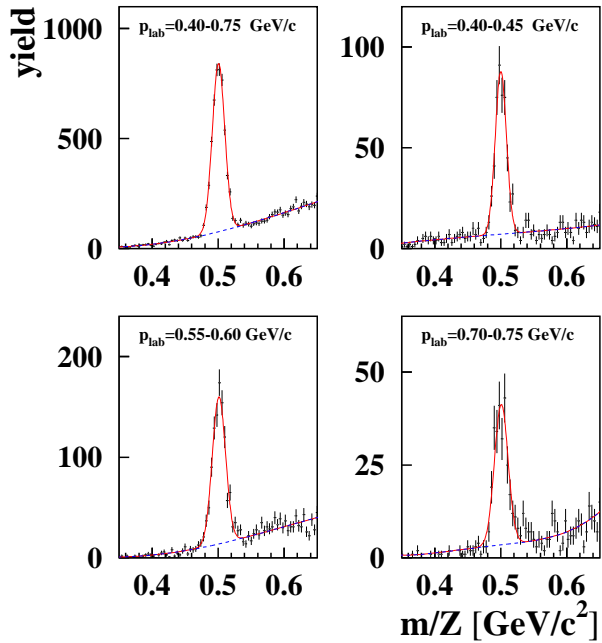,width=9cm}
\caption{(Color online)
  Mass distributions for \mbox{Ni+Ni} collisions at
  $1.1$~\AGeV at $\theta_{\rm lab} = 40^{\circ}$
  for different particle momenta as indicated.
  The solid lines depict the results of a combined
  fit to the background and to the peak at the kaon mass,
  the dashed lines show the background
  part only.
}
\label{mass_fit}
\end{figure}

In order to subtract the remaining background in the mass
distributions, a combined fit using a Gaussian and a polynomial
function to the mass distribution is performed within a window
around the kaon mass. The polynomial part is used to estimate the
background below the kaon peak and is subtracted.
Figure~\ref{mass_fit} shows
mass distributions for \mbox{Ni+Ni} collisions at $1.1$~\AGeV at
$\theta_{\rm lab} = 40^{\circ}$
as an example with a significant background contamination.
The fits are performed separately for each
bin in particle momentum (in most cases having a width of $0.05$~GeV/$c$), 
the upper left
graph shows the distribution integrated
over the full momentum range of this particular
magnetic field setting for illustration purposes only.
The solid lines depict the
result of the combined fit to the background and to the peak at
the kaon mass, the dashed lines show the background part only.

The cross sections are calculated from the number of kaons
$N(p_{\rm lab},\Omega_{\rm lab})$ as
\begin{eqnarray}
   \frac{{\rm d}^2\sigma}{{\rm d}p_{\rm lab} {\rm d}{\mit\Omega}_{\rm lab}}
    & = &
    N(p_{\rm lab},{\mit\Omega}_{\rm lab})
    \cdot \frac{M_{\rm target}}{\rho_{\rm target}
    \cdot d_{\rm target}\cdot N_{\rm A}} \cdot  \nonumber \\
    &   &
    \cdot\frac{1}{N_{\rm proj}}
    \cdot\frac{1}{f_{\rm acc}(p_{\rm lab},{\mit\Omega}_{\rm lab})}
    \cdot\frac{1}{\epsilon(p_{\rm lab})}
\label{Formel:Daten:WQ}
\end{eqnarray}
with $M_{\rm target}$ being the molar mass, $\rho_{\rm target}$
being the density  and $d_{\rm target}$ being the thickness of the
target material. $N_{\rm A}$ denotes Avogadro's constant and
$N_{\rm proj}$ the number of projectiles impinging on the target.

The correction for the geometrical acceptance of the spectrometer
and for the particle decay $f_{\rm acc}(p_{\rm lab},\Omega_{\rm
lab})$ is calculated using a GEANT simulation. The simulated data
sets are analyzed with the same analysis procedure as the
experimental data. The correction is deduced from the ratio
of particles found after the analysis to those initially
simulated.

The total
efficiency $\epsilon(p_{\rm lab})$ is calculated
by multiplying the detector, the DAQ,
and the tracking efficiencies described before
as well as the trigger efficiency $\epsilon_{\rm trigger}(p_{\rm lab})$
and the efficiency of the application of the selection
criteria $\epsilon_{\rm cut}(p_{\rm lab})$.
\begin{figure}
\epsfig{file=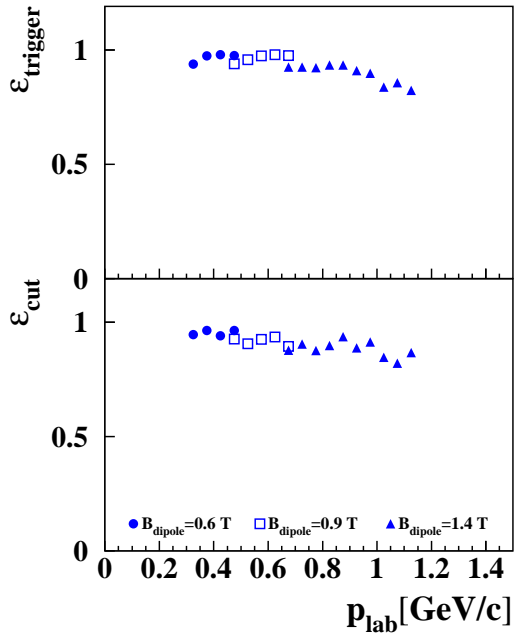,width=8cm}
\caption{(Color online) 
  Trigger efficiencies $\epsilon_{\rm trigger}(p_{\rm lab})$
  for \mbox{Ni+Ni} at $1.1$~\AGeV for the three magnetic
  field settings measured (upper panel),
  and efficiencies of the applied selection criteria
  $\epsilon_{\rm cut}(p_{\rm lab})$ (lower panel) for the same data set.}
\label{efficiencies}
\end{figure}

To determine the efficiency of the trigger system (which is based
on particle velocities) pions and protons measured at magnetic
fields scaled by the mass ratios $m_{\rm pion}/m_{\rm kaon}$ or
$m_{\rm proton}/m_{\rm kaon}$ are used which then have the same
velocities as kaons measured at the nominal fields. These pions
and protons are called ``TOF-simulated kaons'' in the following.
Using an open  trigger condition
that forces every event
for which a particle enters the spectrometer to be recorded
(``w/o TOF'') and
keeping track of the decisions the TOF trigger would have
taken, the efficiency of the TOF trigger can be determined. The
resulting trigger efficiencies are larger than 90\% but they
depend on the particle momentum. The upper panel of
Fig.~\ref{efficiencies} shows $\epsilon_{\rm trigger}(p_{\rm
lab})$ for \mbox{Ni+Ni} at $1.1$~\AGeV for the three magnetic
field settings measured ($|B_{\rm dipole}| = 0.6$, $0.9$, and
$1.4$~T).

The efficiency of the applied selection criteria as a
function of $p_{\rm lab}$ is determined using the background-free
measurements as described above.
In some data sets the statistics of these
settings was not sufficient. In these cases the efficiencies of
the ``geometrical cuts'' were determined using GEANT simulations,
the efficiency of the ``velocity cut'' was determined
using the ``TOF-simulated
kaons''. This method has been validated by comparing its results
to efficiencies calculated from real data in cases where this is
possible. The resulting
efficiencies range from 75\% to 100\%. For the example
\mbox{Ni+Ni} at $1.1$~\AGeV they are depicted in the lower panel
of Fig.~\ref{efficiencies}.

\begin{figure}
\epsfig{file=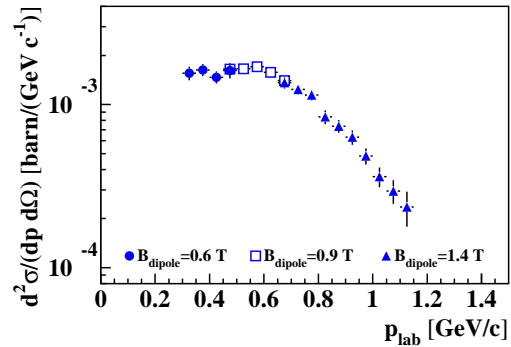,width=8cm}
\caption{(Color online) Differential cross section as a function
  of $p_{\rm lab}$  for \kap in
  \mbox{Ni+Ni} collisions at $1.1$~\AGeV at
  $\theta_{\rm lab} = 40^{\circ}$.
  The different symbols denote data measured at different
  magnetic field settings.}
\label{sig_3fields}
\end{figure}

Figure~\ref{sig_3fields} shows the resulting
differential cross sections for \kap
mesons in \mbox{Ni+Ni} collisions at $1.1$~\AGeV at $\theta_{\rm
lab} = 40^{\circ}$ according to Eq.~(\ref{Formel:Daten:WQ}) as a
function of $p_{\rm lab}$. This
momentum distribution consists of data measured at three different
magnetic field settings ($|B_{\rm dipole}| = 0.6$, $0.9$, and
$1.4$~T) which are analyzed separately but overlap very well.

\section{Experimental Results}
\label{results}

\subsection{Inclusive Reactions}
\label{results_inclusive}

In this section the inclusive cross sections for the production
of \kap and of \kam mesons in \mbox{C+C}, in \mbox{Ni+Ni} and in
\mbox{Au+Au} collisions  are presented as a function of their
laboratory momentum $p_{\rm lab}$ as well as of their energy
$E_{\rm c.m.}$ and their emission angle $\theta_{\rm c.m.}$ in the
center-of-momentum frame. Inclusive means that no centrality
selection has been applied to the data, neither in the analysis
nor implicitly by the experimental setup or the trigger. Let us
recall that the trigger is generated by a time-of-flight signal in
the spectrometer and that the multiplicity detector is not part
of the trigger system. The determination of the functional
dependencies of the production cross sections on $E_{\rm c.m.}$
and on $\theta_{\rm c.m.}$ allows for an extrapolation to phase
space areas not covered by the experiment and for a determination
of integrated production yields. All results of this section are
summarized in Table~\ref{cross_table}. Throughout the figures of
this paper we use a consistent color code for an easy distinction
of the collision systems while of course keeping full descriptions
for black-and-white printed versions: Results of \mbox{Au+Au}
collisions appear in red, \mbox{Ni+Ni} in blue, and \mbox{C+C} in
green.

Some of the data sets presented in this section are published here
for the first time, some have already been
published previously~\cite{laue,laue_epj,menzel,sturm}. As slightly different
procedures have been used in these publications to extrapolate and
to integrate the data, we have recalculated the total
production cross sections using  one consistent procedure for all
data sets.
Earlier measurements with large errors are not taken into
account~\cite{misko,Barth,ahner}.

\subsubsection{Energy Distributions}

\begin{figure}
\vspace{-1cm}
\epsfig{file=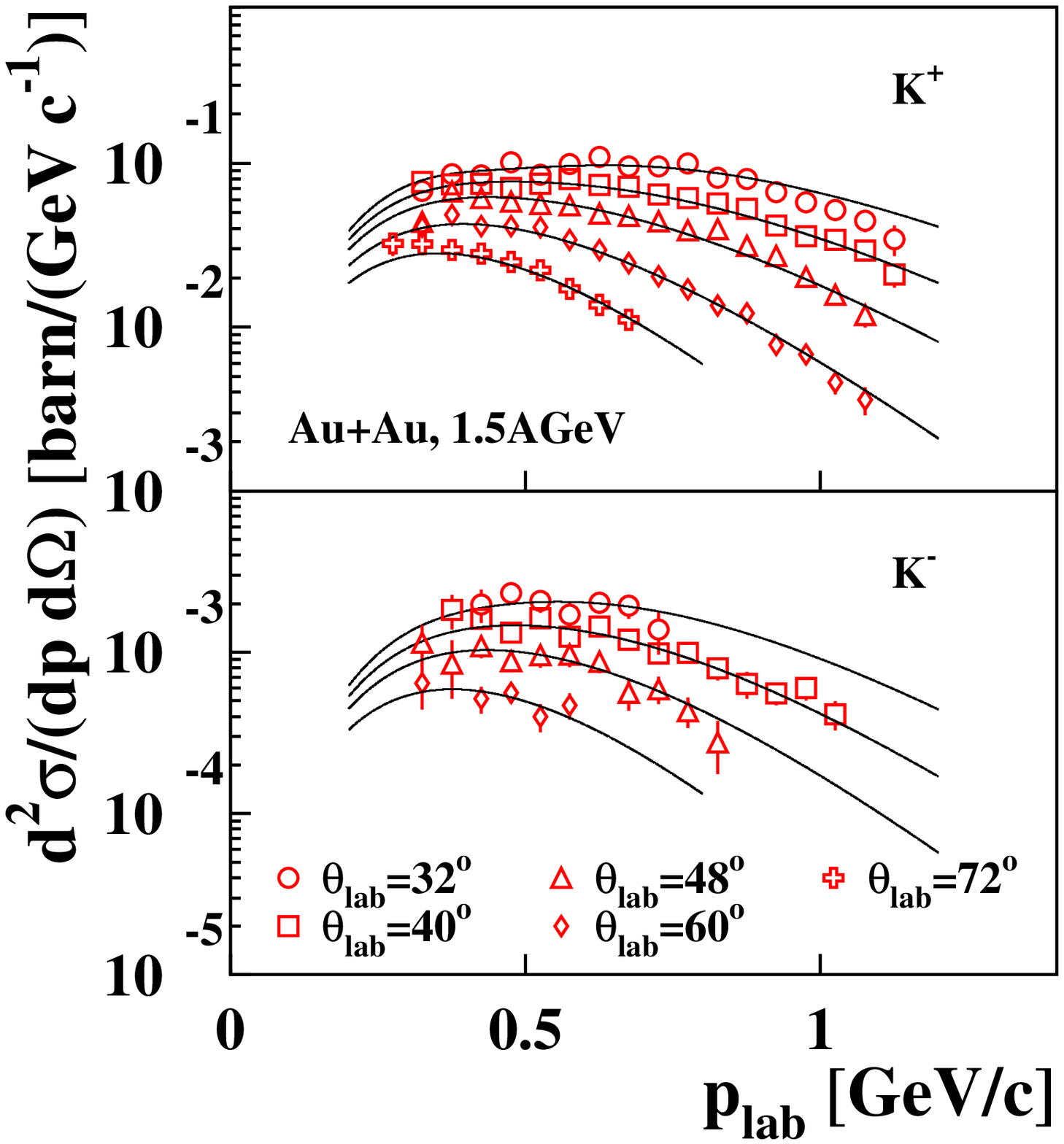,width=7.0cm}
\vspace{-1cm}
\epsfig{file=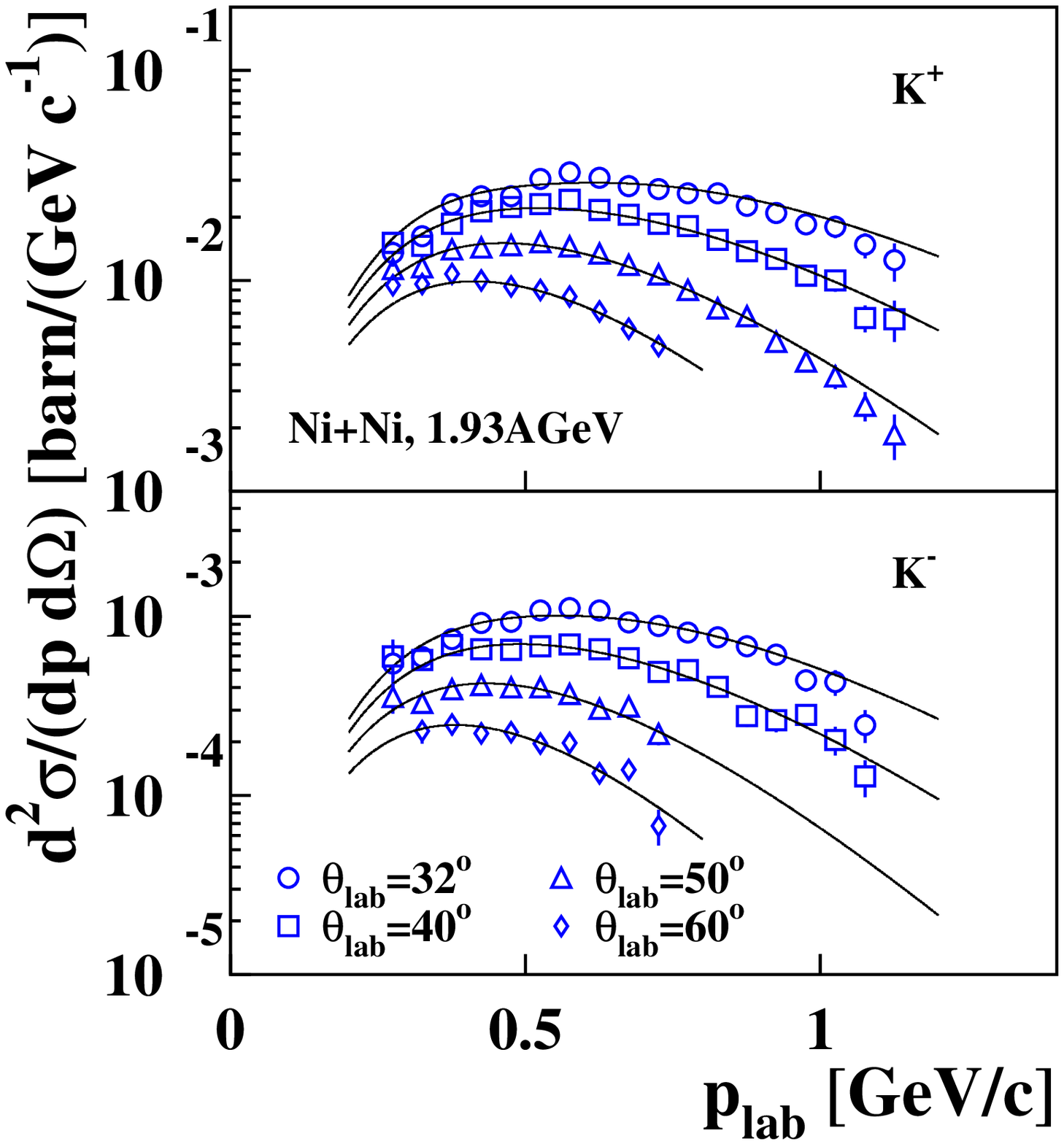,width=7.0cm}
\epsfig{file=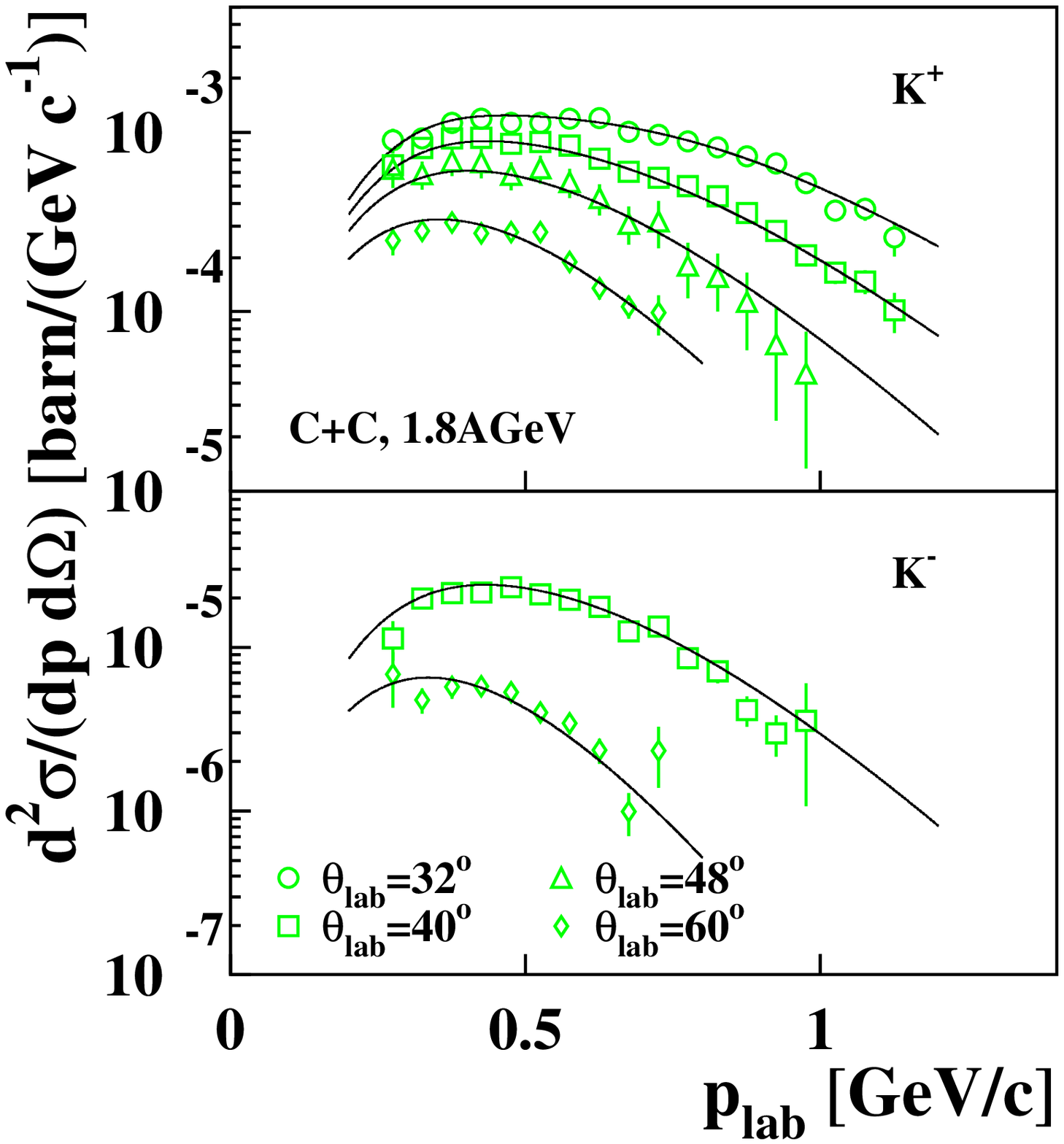,width=7.0cm}
 \caption{(Color online) 
  Inclusive production cross sections for \kap and for \kam
  as a function of the laboratory momentum $p_{\rm lab}$
  for inclusive reactions of \mbox{Au+Au} at $1.5$~\AGeV
  (upper part), of \mbox{Ni+Ni} at $1.93$~\AGeV (middle) and of
  \mbox{C+C} at $1.8$~\AGeV (lower part).
  The lines represent a simultaneous fit to all laboratory angles
  using the distribution according to Eq.~(\ref{eq_3_3}).
}
\label{p_lab}
\end{figure}

Figure \ref{p_lab} shows the inclusive production cross sections
for \kap (upper panels) and for \kam (lower panels) as a
function of their momentum $p_{\rm lab}$ in the laboratory system
for three different collision systems. The upper part of the
figure depicts data taken in \mbox{Au+Au} collisions at a beam
energy of $1.5$~\AGeV[], the middle part those for \mbox{Ni+Ni} at
$1.93$~\AGeV and the lower part those for \mbox{C+C} at
$1.8$~\AGeV[].  To obtain a wide coverage of the phase space
(see Fig.~\ref{pty_kap})
measurements at several polar angle
settings of the spectrometer in the laboratory $\theta_{\rm lab}$
have been performed.
The lines in Fig.~\ref{p_lab} are the
results of simultaneous fits to all angular settings measured for
a given system assuming a Maxwell-Boltzmann shaped dependence of
the invariant production cross section on the center-of-momentum
energy $E_{\rm c.m.}$ and a quadratic dependence on the cosine of
the polar emission angle in the \mbox{c.m.} system
$\theta_{\rm c.m.}$. This procedure will be discussed in detail later.

\begin{figure}
\epsfig{file=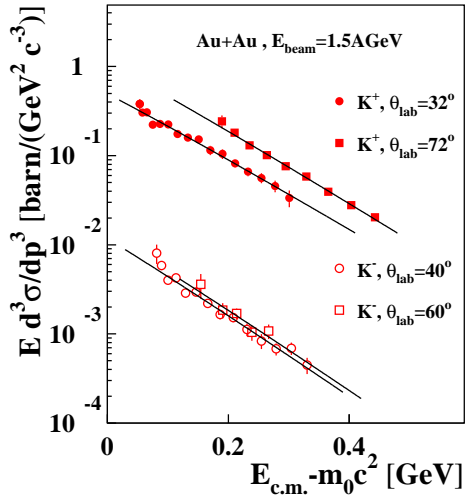,width=7cm}
 \caption{(Color online) Invariant cross sections for \kap (full symbols)
  and for \kam (open symbols) in inclusive
  \mbox{Au+Au} collisions at $1.5$~\AGeV[],
  both at different laboratory angles.
  The lines represent fits according to Eq.~(\ref{eq_3_1}).
  The observation that the data measured at
  different laboratory angles do not coincide indicates a non-isotropic
  polar-angle distribution.}
\label{siginv}
\end{figure}

The necessity for using a non-isotropic distribution in
$\theta_{\rm c.m.}$ is depicted in Fig.~\ref{siginv}. Here the
invariant cross sections $\sigma_{\rm inv} = E \frac{\rm d^{3}
\sigma}{{\rm d} p^{3}}$ are shown as a function of the kinetic
energy in the \mbox{c.m.} system $E_{\rm c.m.} - m_{0}c^{2}$ for
\mbox{Au+Au} at $1.5$~\AGeV[]. Full symbols denote \kap[], open
symbols \kam[]. The data measured at a small angle in the
laboratory system (\kap[]: $\theta_{\rm lab} = 32^{\circ}$,
\kam[]: $\theta_{\rm lab} = 40^{\circ}$) are depicted by circles,
those measured at a large angle (\kap[]: $\theta_{\rm lab} =
72^{\circ}$, \kam[]: $\theta_{\rm lab} = 60^{\circ}$) are represented
by squares. The lines are Maxwell-Boltzmann distributions
\begin{equation}
E  \frac{{\rm d}^{3} \sigma}{{\rm d} p^{3}} \, \sim \, E_{\rm c.m.} \exp{
\left( - \frac{E_{\rm c.m.}}{T} \right) } \label{eq_3_1}
\end{equation}
fitted to the data with $T$ being the inverse slope parameter.
For an isotropic emission in the \mbox{c.m.} system all
spectra of a given particle type are expected to fall on top of
each other regardless of the laboratory angle at
which they have been measured.
For \kap mesons
this is clearly not the case pointing towards an anisotropic
emission.

Since the distributions have been measured at fixed values of
$\theta_{\rm lab}$ in the laboratory, data points at different
particle energies correspond to different emission angles
$\theta_{\rm c.m.}$ in the \mbox{c.m.} system. Therefore, this
anisotropy might affect the determination of the inverse
slope parameter $T$ of the energy spectra if data measured at a
fixed $\theta_{\rm lab}$ would be transformed into the \mbox{c.m.}
system and then fitted.

\begin{figure}
\epsfig{file=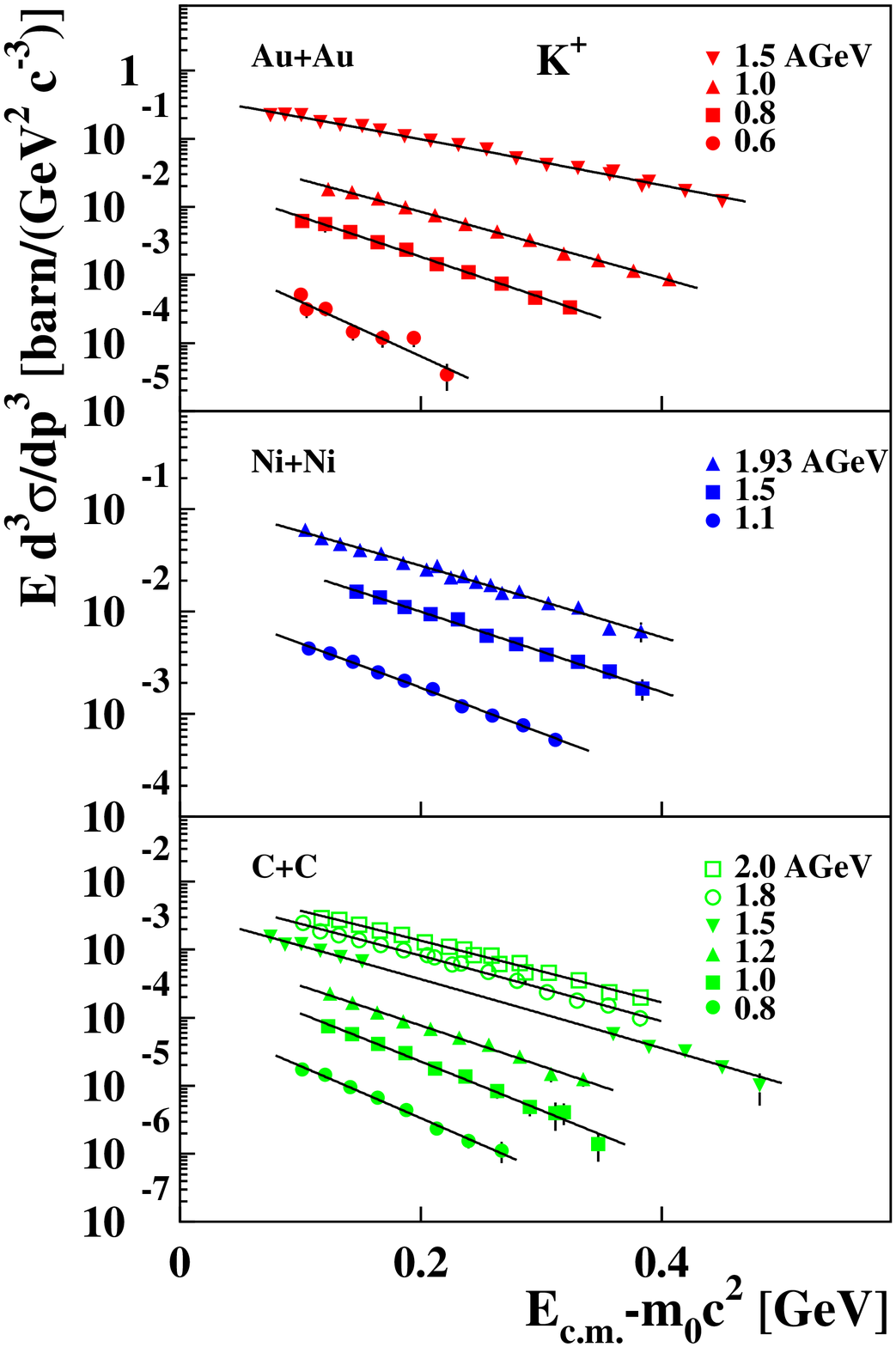,width=7.5cm}
\epsfig{file=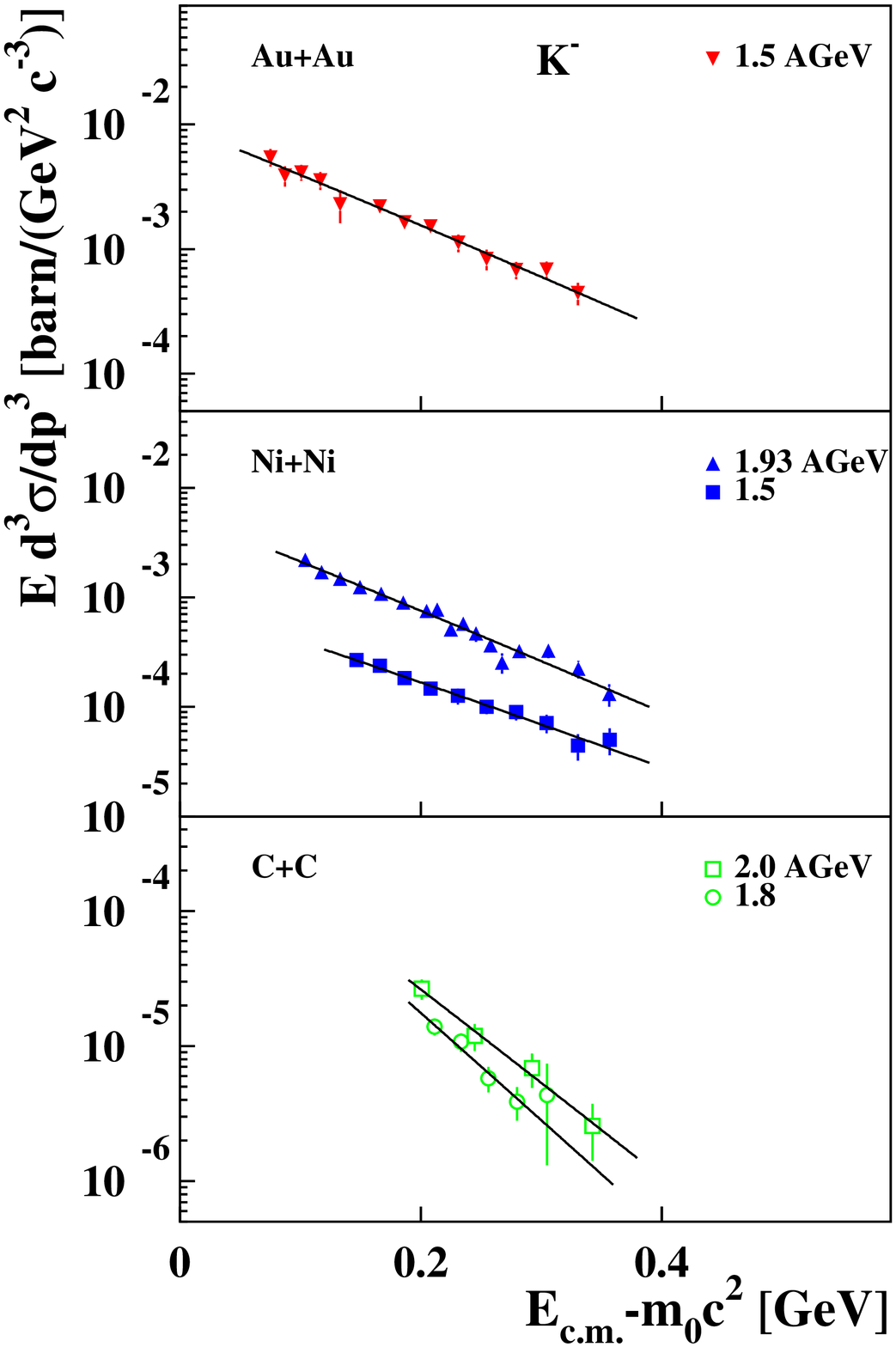,width=7.5cm}
 \caption{(Color online) Inclusive invariant cross sections at midrapidity
 as a function of the kinetic energy $E_{\rm c.m.} - m_{0}c^{2}$
 for \kap (upper part) and for \kam (lower part)
 for the various collision systems and beam energies measured.
 The midrapidity condition is a selection
 of $\theta_{\rm c.m.} = 90^\circ \pm 10^\circ$.}
\label{midrap}
\end{figure}

Therefore, we created ``midrapidity
distributions'' by selecting data points within $\theta_{\rm c.m.} =
90^{\circ} \pm 10^{\circ}$ from measurements at various
laboratory angles.
The results are shown in Fig.~\ref{midrap}, in the upper part for
\kap[], in the lower part for \kam[]. The figure summarizes the
energy distributions of the \kap and of the \kam production as
measured in three different collision systems (\mbox{Au+Au},
\mbox{Ni+Ni}, \mbox{C+C})  at different incident energies
($0.6$~\AGeV up to $2$~\AGeV[]).
The measurements of \kap in \mbox{Au+Au} at $1.135$~\AGeV
and of \kam in \mbox{C+C} at $1.5$~\AGeV are not displayed
in Fig.~\ref{midrap} since they do not cover midrapidity.
The lines are fits to the data
according to Eq.~(\ref{eq_3_1}). The resulting inverse slope
parameters $T_{\rm midrap}$ are given in Table~\ref{cross_table}.

\subsubsection{Polar Angle Distributions}

To extract the angular emission pattern we assume that the
dependence of the invariant cross sections  on
the polar angle $\theta_{\rm c.m.}$ and on the energy $E_{\rm
c.m.}$ can be factorized.
The energy dependence is determined at midrapidity by
fitting Maxwell-Boltzmann distributions to the data as described
above and shown in Fig.~\ref{midrap}. As already mentioned,
each of the data points measured at a given center-of-momentum
energy $E_{\rm c.m.}$ and a laboratory angle $\theta_{\rm lab}$
corresponds to a different emission angle $\theta_{\rm c.m.}$ in
the center-of-momentum frame.

\begin{figure}
\epsfig{file=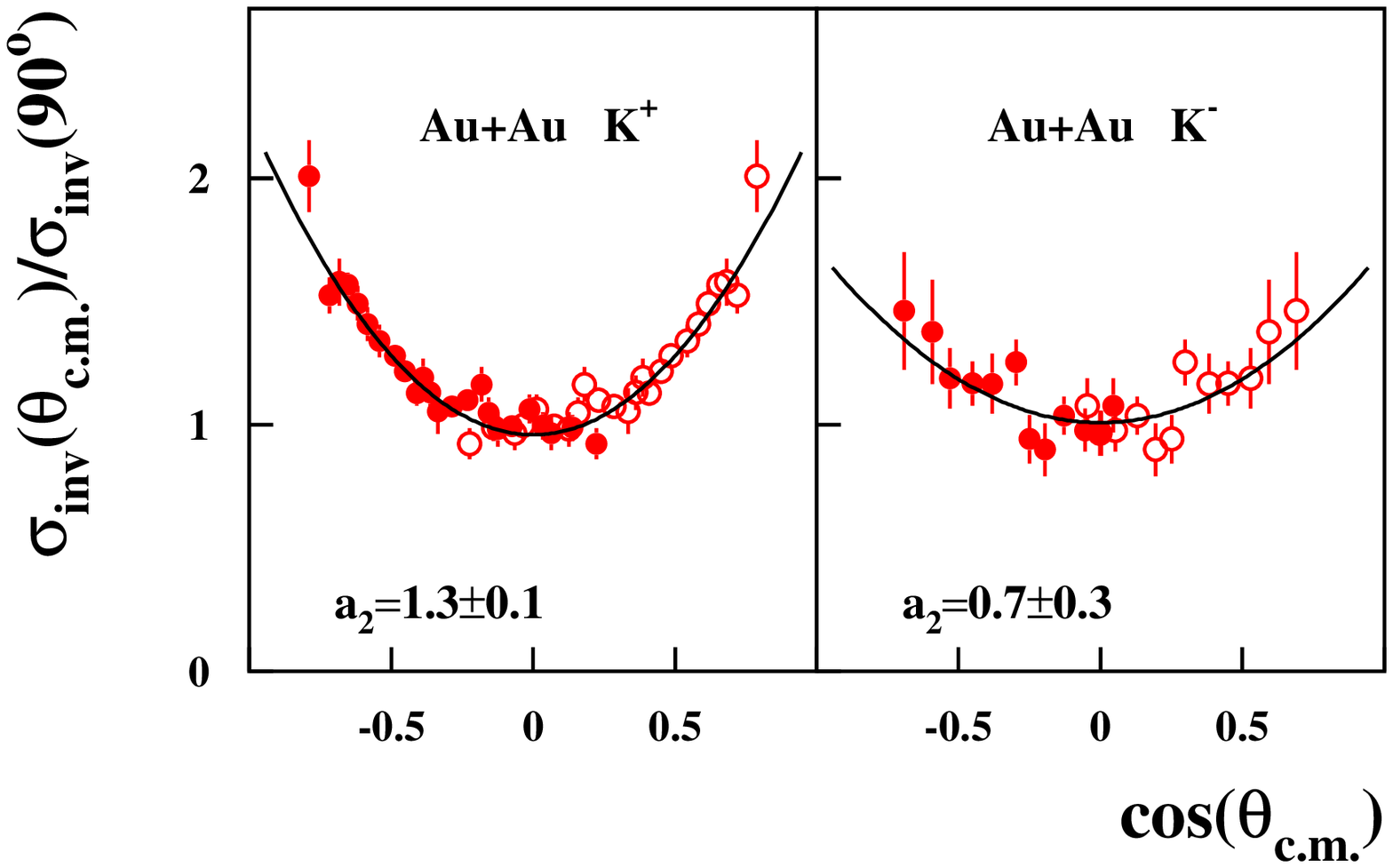,width=8cm}
\epsfig{file=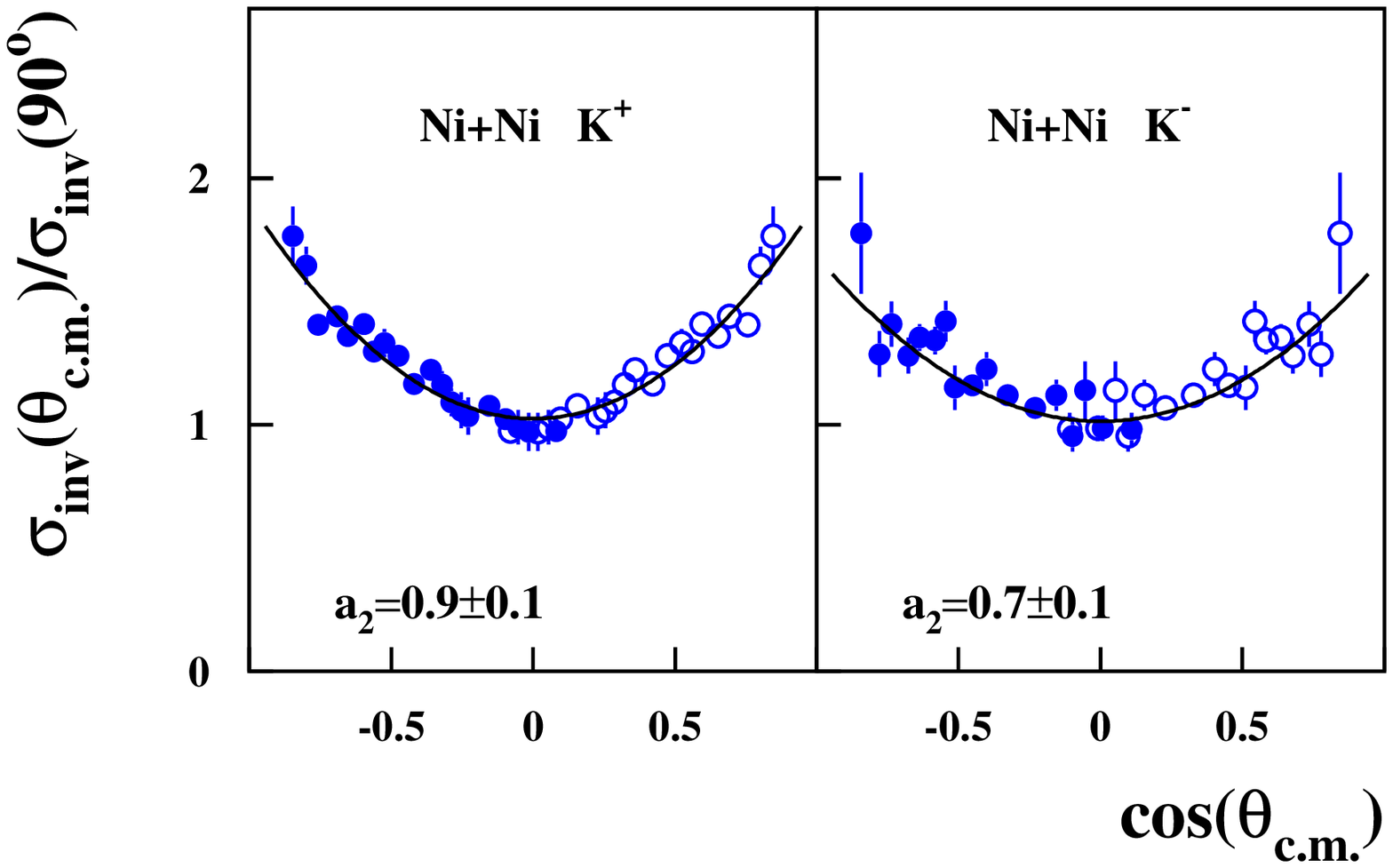,width=8cm}
 \caption{(Color online) Polar angle distributions for inclusive \mbox{Au+Au} collisions at $1.5$~\AGeV (upper
  part) and \mbox{Ni+Ni} at $1.93$~\AGeV (lower part). Full symbols
  denote measured data, open symbols are reflected at
  $\theta_{\rm{c.m.}} = 90^{\circ}$. The lines represent fits according
  to Eq.~(\ref{eq_3_2}).}
\label{ang_dist}
\end{figure}

To disentangle the dependencies on the energy
and on the polar emission angle
we normalized each measured data point $\sigma_{\rm
inv} (E_{\rm c.m.}, \theta_{\rm c.m.})$ to the corresponding
value $\sigma_{\rm inv} (E_{\rm c.m.}, \theta_{\rm
c.m.}=90^{\circ})$. The latter is determined using the fits to
``midrapidity distributions'' according to Eq.~(\ref{eq_3_1}).
Assuming that the energy dependence of the kaon
production is fully described by these midrapidity fits the results
are the polar-angle emission pattern.
They are shown in Fig.~\ref{ang_dist}
as a function of
$\cos(\theta_{\rm c.m.})$,
in the upper part
for \mbox{Au+Au} at $1.5$~\AGeV and in the lower part
for \mbox{Ni+Ni} at $1.93$~\AGeV[].
Full symbols are measured data
points, open symbols have been reflected at $\theta_{\rm c.m.} =
90^{\circ}$ since
for mass-symmetric systems the
polar angle distributions have to be symmetric around $\theta_{\rm
c.m.}=90^{\circ}$.
Both systems show a forward-backward preference in
the emission pattern which is more pronounced for \kap (left hand
side) than for \kam (right hand side).

To quantify the anisotropy, the distributions have been fitted
with a quadratic dependence on $\cos(\theta_{\rm c.m.})$
\begin{equation}
\frac{{\rm d} \sigma}{{\rm d} \cos(\theta_{\rm c.m.})} \, \sim \, 1 \, + \,
a_{2}^{\rm div} \cos^{2}(\theta_{\rm c.m.}) \label{eq_3_2}
\end{equation}
as depicted by the lines in the figure. This procedure has been
applied to most data sets resulting in the values for
$a_{2}^{\rm div}$ as given in Table~\ref{cross_table}.
In the cases for which only one laboratory angle has been measured
the coverage in $\theta_{\rm c.m.}$ is rather small
making the determination of $a_{2}^{\rm div}$ impossible.

The super(sub)scripts {\rm 'div'} for the angular anisotropy and
{\rm 'midrap'} for the inverse
slope parameters are used for the two-step procedure
presented above.
Since the energy dependence of the kaon production is well described
by a Maxwell-Boltzmann distribution and the polar angle
distribution by a quadratic dependence on $\cos(\theta_{\rm c.m.})$
we additionally performed simultaneous fits
to all momentum distributions measured at different laboratory
angles $\theta_{\rm lab}$ for a given system using the combined
function
\begin{equation}
E \frac{{\rm d}^{3} \sigma}{{\rm d} p^{3}} \, = \, C \left[1  + a_{2}^{\rm sf}
\cos^{2}(\theta_{\rm c.m.}) \right] E_{\rm c.m.}  \exp{ \left(
- \frac{E_{\rm c.m.}}{T_{\rm sf}} \right) } \label{eq_3_3}
\end{equation}
with $a_{2}^{\rm sf}$, $T_{\rm sf}$ and the normalization $C$
being the three variable parameters.
The results of this procedure are denoted by the
super(sub)script {\rm 'sf'}. For \mbox{Au+Au} at
$1.5$~\AGeV[], \mbox{Ni+Ni} at $1.93$~\AGeV[], and \mbox{C+C} at
$1.8$~\AGeV the results of these fits are shown as solid lines in
Fig.~\ref{p_lab}. The parameters obtained for all collision
systems at all beam energies are given in Table~\ref{cross_table}.
They agree very well with the values obtained by the two-step
procedure denoted by $T_{\rm midrap}$ and by $a_2^{\rm div}$. 
The combined fits in addition
provide the correlations between the three
parameters and thus the full error matrix which is necessary for
calculating the errors of the integrated production cross sections.
In those cases for which only one angle has been measured we take an
interval for the polar angle anisotropy $a_{2}^{\rm sf}$, denoted
by square brackets in Table~\ref{cross_table}, with values set
according to the trend at neighboring beam energies. This
variation of $a_{2}^{\rm sf}$ yields additional errors on the
inverse slope parameters $T_{\rm sf}$ as well as on the integrated cross
sections as tagged by the superscript {\rm 'a$_2$'} in
Table~\ref{cross_table}.

\subsubsection{Total Production Cross Sections}

The results of the simultaneous fits have been used to extrapolate
the data to phase-space regions not covered by the experiment and
to calculate total production cross sections by integrating
Eq.~(\ref{eq_3_3}) over the full phase space. 
The extrapolation in
$E_{\rm c.m.}$ contributes with about $35\%$ to the total
production cross sections.
The resulting total production cross sections for \kap
and for \kam for all collision systems and for all beam energies
are summarized in Table~\ref{cross_table}. The error bars of the data
points in the figures showing energy spectra and polar angle distributions
contain the statistical uncertainties as well as point to point systematic
errors due to the background subtraction.
The overall systematic error of the absolute normalization
is stated separately in Table~\ref{cross_table} and is quadratically 
added to the statistical errors in all figures comparing 
cross section or multiplicities from different collision systems.

In several cases only one polar angle has been measured. As
already described for the inverse slope parameters $T$ in the
previous section, for each of those cases an interval for $a_2$
guided by the systematics given by neighboring beam energies has
been used to determine an additional error on the integrated cross
section $\sigma$ denoted by the superscript 'a$_{\rm 2}$' in
Table~\ref{cross_table}.

Part of the data has been published earlier with the methods to
extrapolate the measured data to the full phase space being
slightly different in the various publications. In this paper, we
consistently apply one single method to extrapolate and
integrate the data.

For \mbox{C+C} collisions $\sigma$ and $T$ for \kap and for \kam
were published~\cite{laue} assuming
the angular anisotropy for all
incident energies to contribute with 20\% to
the total cross section independent of the incident
energy which is equivalent to $a_2 = 0.6$.
Now we determine $a_2$ for each measurement separately.
The differences between
the values for $\sigma$ and $T$ in Table~\ref{cross_table} and
those published in \cite{laue} are nevertheless
smaller then the statistical errors.

In the case of \mbox{Ni+Ni}, cross sections for the
\kap and for the \kam production at $1.93$~\AGeV have been
published in \cite{menzel}
with the angular anisotropy being taken from a two-step
procedure rather than from a simultaneous fit.
Also in this case the differences
between the results in Table~\ref{cross_table} and the previously
published values are smaller than the statistical errors.

For \mbox{Au+Au} $\sigma$, $T$ and $a_2$ have been published for
\kap[]~\cite{sturm}. The results in Table~\ref{cross_table} for
$0.6$, $0.8$, $1.0$ and $1.135$~\AGeV differ less than the
statistical errors from the published values.
For \kap at $1.5$~\AGeV the results published in
\cite{sturm} are: $\sigma=267\pm30$~mb,
$T=100\pm5$~MeV and $a_2=1.06\pm0.3$. They correspond to a low
statistics measurement using a thicker target reducing the
effective beam energy for the \kap production to $1.46$~\AGeV[].
This difference in effective energy accounts for a difference
of about $10\%$ in cross section compared to the new
high statistics experiment reported upon in this paper using a
thinner target (with $E_{eff}=1.48$~\AGeV[]). 
In addition in the experiment described in
\cite{sturm} the anisotropy in the polar angle emission pattern
was underestimated due to a reduced coverage in $\theta_{\rm lab}$
accounting for an additional $5\%$ difference in the cross section.

\begin{figure}
\epsfig{file=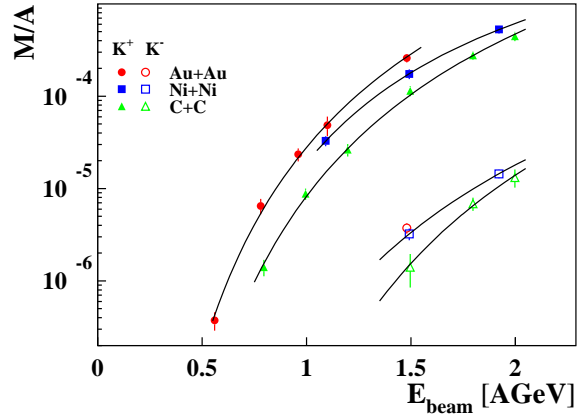,width=8.5cm}
 \caption{(Color online) 
    Multiplicities of \kap (full symbols) and of \kam mesons
    (open symbols) per mass number $A$
    of the respective collision system as a
    function of the beam energy. The lines represent fits
    to the data according to Eq.~(\ref{eq_kolomeitsev}).}
\label{k_sigma_Energy}
\end{figure}

To calculate particle multiplicities $M$ the integrated
production cross sections $\sigma$ (see Table~\ref{cross_table})
need to be divided by the total reaction cross section
$\sigma_r$ which cannot be determined easily as the
particle-multiplicity distribution measured with a multiplicity
trigger condition has a cutoff at low multiplicities (see also
Sect.~\ref{results_centrality}). Therefore, we determine the
reaction cross sections using Glauber
calculations~\cite{glauber} resulting in $\sigma_r(\rm Au+Au) =
6.8$~barn, $\sigma_r(\rm Ni+Ni) = 3.1$~barn, and $\sigma_r(\rm
C+C) = 0.95$~barn. Figure~\ref{k_sigma_Energy} summarizes the
multiplicities of \kap and of \kam mesons as a function of the
beam energy as determined in inclusive reactions (\mbox{Au+Au},
\mbox{Ni+Ni}, \mbox{C+C}) and normalized to the mass number $A$ of
the respective colliding nuclei.  Both particle species exhibit
strongly rising excitation functions as expected due to the
proximity of the thresholds in binary NN collisions (1.58~GeV
for \kap[], 2.5~GeV for \kam[]). The solid lines reflect fits
according to the formula~\cite{Kolo}
\begin{equation}
\frac{M_{\rm K}}{A} \, = \, C \sqrt{T_{\rm max}} \,
\,\exp \left[ {-\frac{E_{\rm thr}}{T_{\rm max}}} \right] \, ,
\label{eq_kolomeitsev}
\end{equation}
with $T_{\rm max} \,   = \, T_0 \cdot (E_{\rm beam})^\eta$. The
variable parameters in the fit are $C$, $T_0$ and $\eta$.
 As well for \kap as for \kam the
multiplicities per mass number $A$ increase with system size
from \mbox{C+C} to \mbox{Au+Au} at the same incident energy.

\begin{table*}[ht]
\begin{center}
\begin{tabular}{ c c c c c c c c c c l } \hline
reaction  & $E_{\rm beam}$ & $\theta_{\rm lab}$ &$\sigma$ 
& $T_{\rm sf}$ & $a_2^{\rm sf}$& $T_{\rm midrap}$& $a_2^{\rm div}$& \\
& [$A$~GeV] & [$^\circ$]     & [mb]    & [MeV]     &            & [MeV]       &      &  \\
\hline
{\bf K}$^{\boldsymbol{+}}$ &                       &                                       &                                  &                  &             &                & \\
C+C   & 0.8   & 44             & $0.016\pm0.002^{\rm stat}\pm0.0016^{\rm syst}\pm0.002^{\rm a_2}$         & $54\pm4\pm1^{\rm a_2}$               & $[-0.5,0.5]$      & $52 \pm 4$  & ------         & \\
      & 1.0   & 44,54,70       & $0.1\pm0.01^{\rm stat}\pm0.01^{\rm syst}$                          & $60\pm3$                         & $0.25\pm 0.27$   & $56 \pm 4$  & $0.18\pm 0.27$ & \\
      & 1.2   & 40             & $0.3\pm0.03^{\rm stat}\pm0.03^{\rm syst}\pm0.02^{\rm a_2}$            & $67\pm4\pm2^{\rm a_2}$               & $[0.0,1.0]$    & $67 \pm 5$  & ------         & \\
      & 1.5   & 32,48          & $1.3\pm0.1^{\rm stat}\pm0.13^{\rm syst}$                           & $77\pm3$                         & $0.67\pm 0.30$   & $77 \pm 5$  & $0.71\pm 0.23$ & \\
      & 1.8   & 32,40,48,60    & $3.15\pm0.13^{\rm stat}\pm0.32^{\rm syst}$                         & $81\pm2$                         & $1.21\pm0.15$    & $81 \pm 3$  & $1.25\pm0.14$  & \\
      & 2.0   & 32,40          & $5.1\pm0.3^{\rm stat}\pm0.5^{\rm syst}$                           & $86\pm3$                         & $1.13\pm0.23$    & $85 \pm 5$  & $1.12\pm0.17$  & \\
\hline
Ni+Ni & 1.1   & 40             & $5.9\pm0.3^{\rm stat}\pm0.6^{\rm syst}\pm0.3^{\rm a_2}$               & $87\pm3\pm1^{\rm a_2}$               & $[0.5,1.0]$      & $87\pm  4$  & ------         & \\
      & 1.5   & 40             & $31.4\pm1.3^{\rm stat}\pm3.1^{\rm syst}\pm1.7^{\rm a_2}$              & $101\pm3\pm3^{\rm a_2}$              & $[0.5,1.0]$      & $97\pm  7$  & ------         & \\
      & 1.93  & 32,40,50,60    & $95\pm2^{\rm stat}\pm9.5^{\rm syst}$                              & $112\pm2$                        & $0.90\pm0.06$    & $108\pm 5$  & $0.85\pm 0.06$ & \\
\hline
Au+Au & 0.6   & 50             & $0.5\pm0.1^{\rm stat}\pm0.05^{\rm syst}\pm0.03^{\rm a_2}$              & $49\pm6\pm1^{\rm a_2}$               & $[0.75,1.25]$    & $50\pm 7$   & ------         & \\
      & 0.8   & 44,84          & $8.7\pm1.4^{\rm stat}\pm0.9^{\rm syst}$                           & $67\pm4$                         & $1.20\pm 0.42$   & $66\pm 5$   & $1.25\pm 0.4$  & \\
      & 1.0   & 44,84          & $31.4\pm4^{\rm stat}\pm3.1^{\rm syst}$                            & $82\pm3$                         & $1.07\pm 0.24$   & $81\pm 6$   & $1.16\pm 0.2$ & \\
      & 1.135 & 56             & $65\pm14^{\rm stat}\pm6.5^{\rm syst}\pm2^{\rm a_2}$                   & $89\pm9\pm1^{\rm a_2}$               & $[0.75,1.25]$    & ------      & ------         & \\
      & 1.5   & 32,40,48,60,72 & $346\pm9^{\rm stat}\pm35^{\rm syst}$                             & $111\pm2$                        & $1.25\pm 0.09$   & $111\pm 5$  & $1.28\pm 0.09$ & \\
\hline \hline
{\bf K}$^{\boldsymbol{-}}$ &       &                &                                       &                                  &                  &             &                & \\
C+C   & 1.5   & 40             & $0.016\pm0.006^{\rm stat}\pm0.0016^{\rm syst}\pm0.001^{\rm a_2}$         & $50\pm12\pm7^{\rm a_2}$              & $[0.0,1.0]$      & ------      & ------         & \\
      & 1.8   & 40,60          & $0.078\pm 0.01^{\rm stat}\pm0.008^{\rm syst}$                       & $65\pm8$                         & $0.80\pm0.25$    & $55\pm8$    & ------         & \\
      & 2.0   & 40             & $0.15\pm0.03^{\rm stat}\pm0.015^{\rm syst}\pm0.004^{\rm a_2}$           & $57\pm7\pm2^{\rm a_2}$               & $[0.5,1.5]$      & $58\pm9$    & ------         & \\
\hline
Ni+Ni & 1.5   & 40             & $0.58\pm0.05^{\rm stat}\pm0.06^{\rm syst}\pm0.03^{\rm a_2}$            & $89\pm6\pm2^{\rm a_2}$               & $[0.5,1.0]$      & $98\pm 10$  & ------         & \\
      & 1.93  & 32,40,50,60    & $2.6\pm0.1^{\rm stat}\pm0.26^{\rm syst}$                           & $89\pm2$                         & $0.70\pm 0.09$   & $84\pm 4$   & $0.66\pm 0.09$ & \\
\hline
Au+Au &  1.5  & 32,40,48,60    & $5.0\pm 0.4^{\rm stat}\pm0.5^{\rm syst}$                          & $87\pm 4$                        & $0.56\pm 0.26$     & $91\pm 8$  & $0.64\pm 0.26$   & \\
\hline
\end{tabular}
\caption[]{Summary of the integrated production cross sections of
\kap and of \kam mesons. The inverse slope parameters $T_{\rm sf}$
and $T_{\rm midrap}$ as well as the angular anisotropies $a_2^{\rm sf}$ and
$a_2^{\rm div}$ have been determined by two different procedures as
explained in the text. 
For systems with only one measured angle $\theta_{\rm lab}$ 
the angular anisotropies $a_2^{\rm sf}$ have not been fitted but 
set to an interval guided by neighbouring beam energies. 
It is given in square brackets. The additional errors
for $\sigma$ and $T_{\rm sim}$ due to this procedure
are denoted by the superscript {\rm 'a$_2$'}.} 
\label{cross_table}
\end{center}
\end{table*}

\subsection{Centrality Dependence}
\label{results_centrality}

\begin{figure}
\epsfig{file=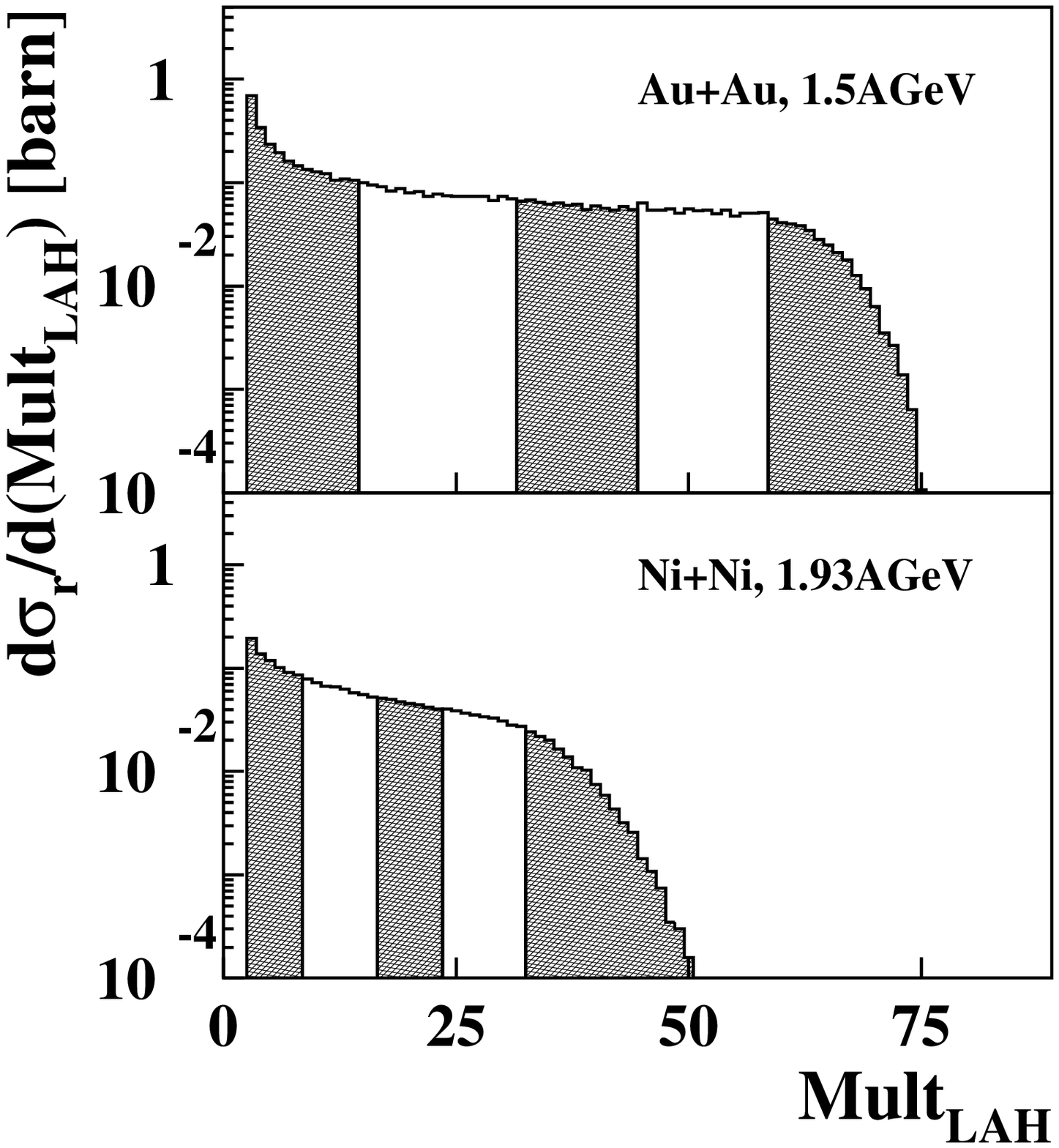,width=7.5cm}
 \caption{ 
  The reaction cross section as a function of the charged
  particle multiplicity in the
  Large-Angle Hodoscope
  for \mbox{Au+Au} collisions at $1.5$~\AGeV (upper panel)
  and for \mbox{Ni+Ni} at $1.93$~\AGeV (lower panel).
  The shaded areas denote the
  five centrality bins.}
\label{mul_dis}
\end{figure}

The collision centrality was derived from the multiplicity of
charged particles measured in the Large-Angle Hodoscope ($Mult_{\rm LAH}$).
Figure \ref{mul_dis} shows the respective multiplicity
distributions for \mbox{Au+Au} at $1.5$~\AGeV
(upper panel) and for \mbox{Ni+Ni} at $1.93$~\AGeV (lower panel)
measured with a multiplicity trigger.
In order to study the centrality dependence of the \kap and of the \kam
production the data were grouped into five centrality bins both
for \mbox{Ni+Ni} at $1.5$ and at $1.93$~\AGeV as well as for
\mbox{Au+Au} at $1.5$~\AGeV[]. These bins are also depicted in
Fig.~\ref{mul_dis}.
The distributions have been 
normalized to the beam intensity, to the target thickness and to
the efficiency of the DAQ system so that the area between the
respective bin boundaries represents the corresponding 
fraction of the total
reaction cross sections $\sigma_r$ for a given bin. Very
peripheral collisions in the first centrality bin might be
missed by the multiplicity trigger,
however, kaons from such peripheral events are measured since
they are triggered by the time-of-flight trigger in the
spectrometer. The fraction of $\sigma_r$ for this most peripheral
bin is determined by taking the inclusive total reaction cross
section from a Glauber calculation~\cite{glauber}
and subtracting the sum of the experimentally measured values of
the four other centrality bins. For \mbox{Au+Au}, the five bins
correspond to $0 \, - \, 5.4\%$, $5.4\% \, - \, 18.1\%$, $18.1\%
\, - \, 31.1\%$, $31.1\% \, - \, 52.3\%$, and  $52.3\% \, - \,
100\%$ of $\sigma_r$ from central to peripheral collisions, for
\mbox{Ni+Ni} to  $0 \, - \, 4.4\%$, $4.4\% \, - \, 15.0\%$,
$15.0\% \, - \, 26.5\%$, $26.5\% \, - \, 45.9\%$, and  $45.9\% \,
- \, 100.0\%$ of $\sigma_r$. The corresponding
mean numbers of participating nucleons $A_{\rm
part}$ have as well been calculated using Glauber
calculations.

\subsubsection{Multiplicities}

\begin{figure}
\epsfig{file=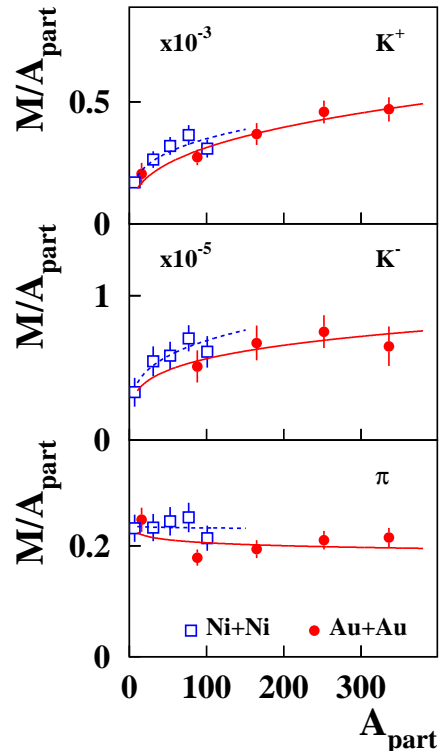,width=7cm}
 \caption{(Color online) 
  Dependence of the multiplicities of \kap (upper panel)
  and of \kam mesons (middle panel)
  as well as of pions (lower panel) on \Apart[]. Full symbols denote
  \mbox{Au+Au}, open
  symbols \mbox{Ni+Ni}, both at $1.5$~\AGeV[]. The lines are functions
  $M \sim A^{\alpha}_{\rm part}$ fitted to the data separately for
  \mbox{Au+Au} (solid lines) and \mbox{Ni+Ni} (dashed lines).
  The data have been measured at $\theta_{\rm{lab}} = 40^{\circ}$.}
\label{Au_ratio}
\end{figure}

The multiplicity of a particle species for each
centrality bin is defined as 
$M = \sigma/(f\cdot\sigma_{\rm r})$ with $\sigma$
being the production cross section for the respective
particle species and $(f\cdot\sigma_{\rm r})$ being the fraction of the
total reaction cross-section for the respective bin. Figure
\ref{Au_ratio} presents the multiplicities per number of
participating nucleons $M/A_{\rm part}$ as a function of $A_{\rm
part}$ at a beam energy of $1.5$~\AGeV for \mbox{Au+Au}
and for \mbox{Ni+Ni}. To calculate these
multiplicities only data measured at $\theta_{\rm lab} = 40^{\circ}$
have been used since for \mbox{Ni+Ni} this is the only laboratory
angle measured at that beam energy.
The \kap are shown in the upper panel, the \kam in
the middle panel and the pions in the lower panel. Since neutral
particles can not be detected by the spectrometer, the total pion
multiplicity has been calculated as $M(\pi) \, = \, 3/2 \, M(\pi^+)
\, + \, 3/2 \, M(\pi^-)$. In all panels full symbols denote
\mbox{Au+Au}, open symbols \mbox{Ni+Ni}.

The lines in Fig.~\ref{Au_ratio} are functions
$M \sim A^{\alpha}_{\rm part}$ fitted to the data
separately for \mbox{Au+Au} (continuous lines) and for \mbox{Ni+Ni}
(dashed lines).
For both systems the pions show a linear dependence of $M$ on
\Apart with $\alpha$ close to 1 ($\alpha_{\pi}(\rm{Au}) = 0.96
\pm 0.05$, $\alpha_{\pi}(\rm{Ni}) = 1.0 \pm 0.05$) which means that the
number of pions produced is proportional to the number of nucleons
participating in the reaction. For \kap on the other hand
the multiplicities rise stronger then linear with $A_{\rm
part}$ ($\alpha_{\rm K^+}(\rm{Au}) =
1.34 \pm 0.16$, $\alpha_{\rm K^+}(\rm{Ni}) = 1.26 \pm 0.06$).
The same holds for \kam
($\alpha_{\rm K^-}(\rm{Au}) = 1.22 \pm 0.27$, $\alpha_{\rm K^-}(\rm{Ni}) = 1.25 \pm
0.12$).

Both, the \kap and the \kam
multiplicities rise similarly with centrality leading to a
nearly constant \kam/\kap ratio as a function of \Apart
(see Fig.~\ref{kmkp_apart} in
Sect.~\ref{discussion_coupling_kmkp}) although their production
thresholds differ significantly. This will be discussed in detail
in Sect.~\ref{discussion_coupling_kmkp}.

\subsubsection{Energy Distributions}

Although the multiplicities of
\kap  and of \kam mesons per \Apart
in \mbox{Au+Au} and \mbox{Ni+Ni} show the same rise
with $A_{\rm{part}}$ and even have the same absolute values,
significant differences between \kap and \kam have been found~\cite{AF}.

\begin{figure}
\epsfig{file=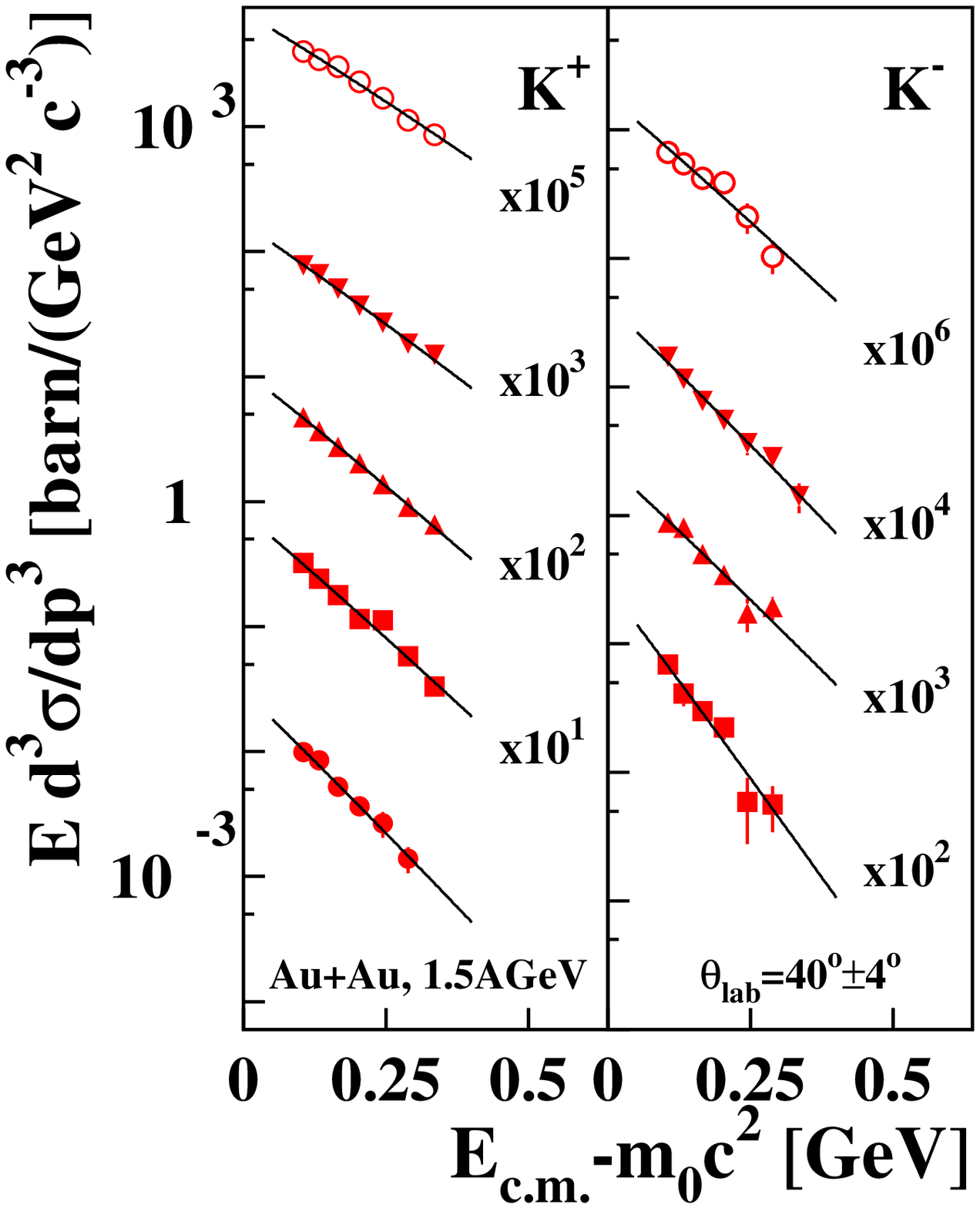,width=8cm}
\epsfig{file=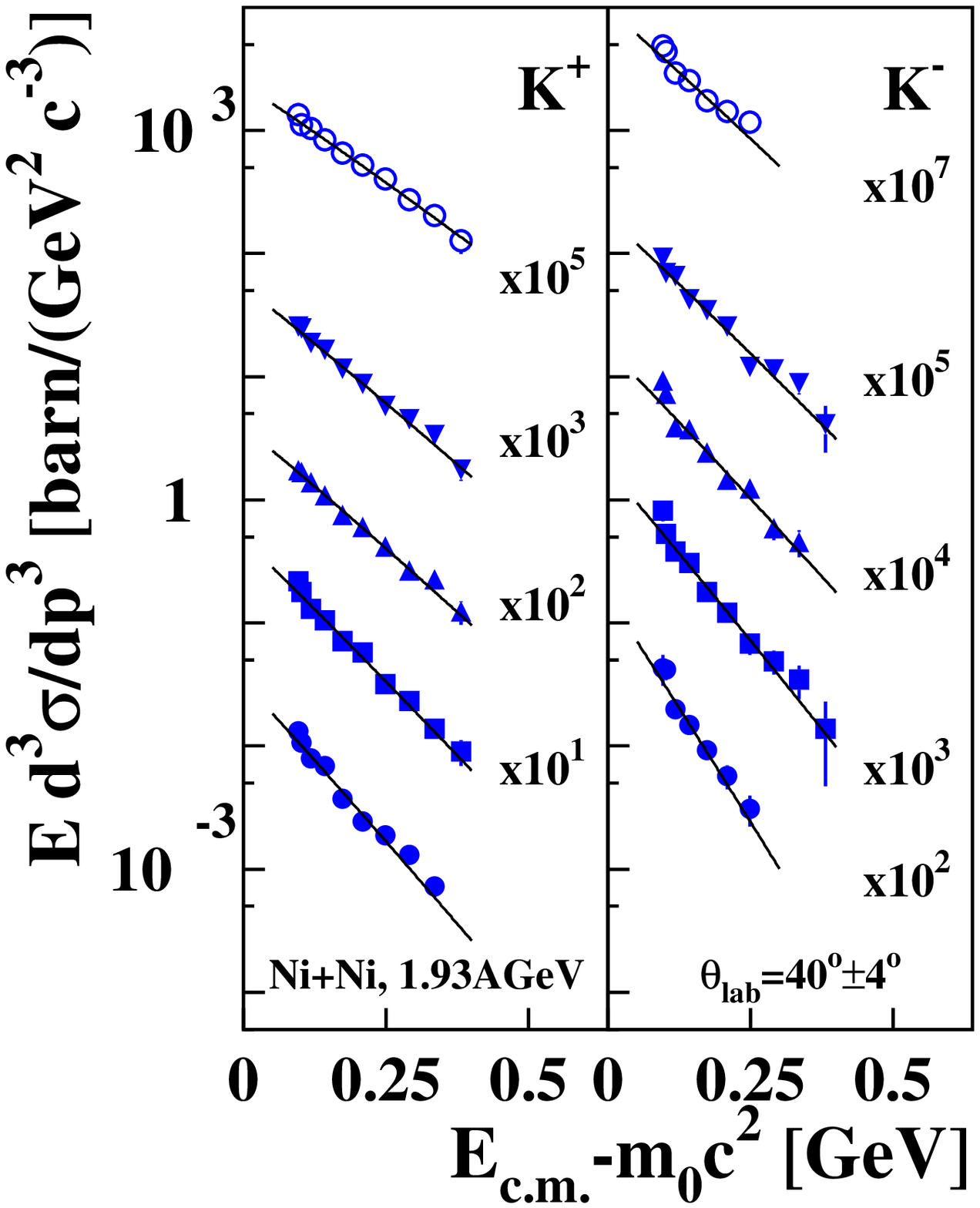,width=8cm}
\caption{(Color online) 
  Energy distributions of the invariant cross sections for
  \mbox{Au+Au} at $1.5$~\AGeV (upper part) and for
  \mbox{Ni+Ni} at $1.93$~\AGeV
  (lower part)
  close to midrapidity for different centralities. The uppermost spectra
  correspond to the most central collisions.
  The subsequent bins are shown from
  top to bottom with decreasing centrality. The lines represent fits to
  the data according to Eq.~(\ref{eq_3_1}).}
\label{spec_cm}
\end{figure}

Figure \ref{spec_cm} shows the invariant cross sections
for \kap and for \kam
mesons measured close to midrapidity ($\theta_{\rm lab}=40^{\circ}$)
as a function of the kinetic
energy in the center-of-momentum system, in the upper part for
\mbox{Au+Au} collisions at $1.5$~\AGeV[], in the lower part for
\mbox{Ni+Ni} at $1.93$~\AGeV[]. The left panel of each graph
depicts \kap[], the right panel \kam[]. The uppermost distributions
correspond to the most central reactions, the subsequent bins are
shown from the top to the bottom of each graph with decreasing
centrality. The solid lines represent Maxwell-Boltzmann
distributions according to Eq.~(\ref{eq_3_1}) fitted to the data.

\begin{figure}
\epsfig{file=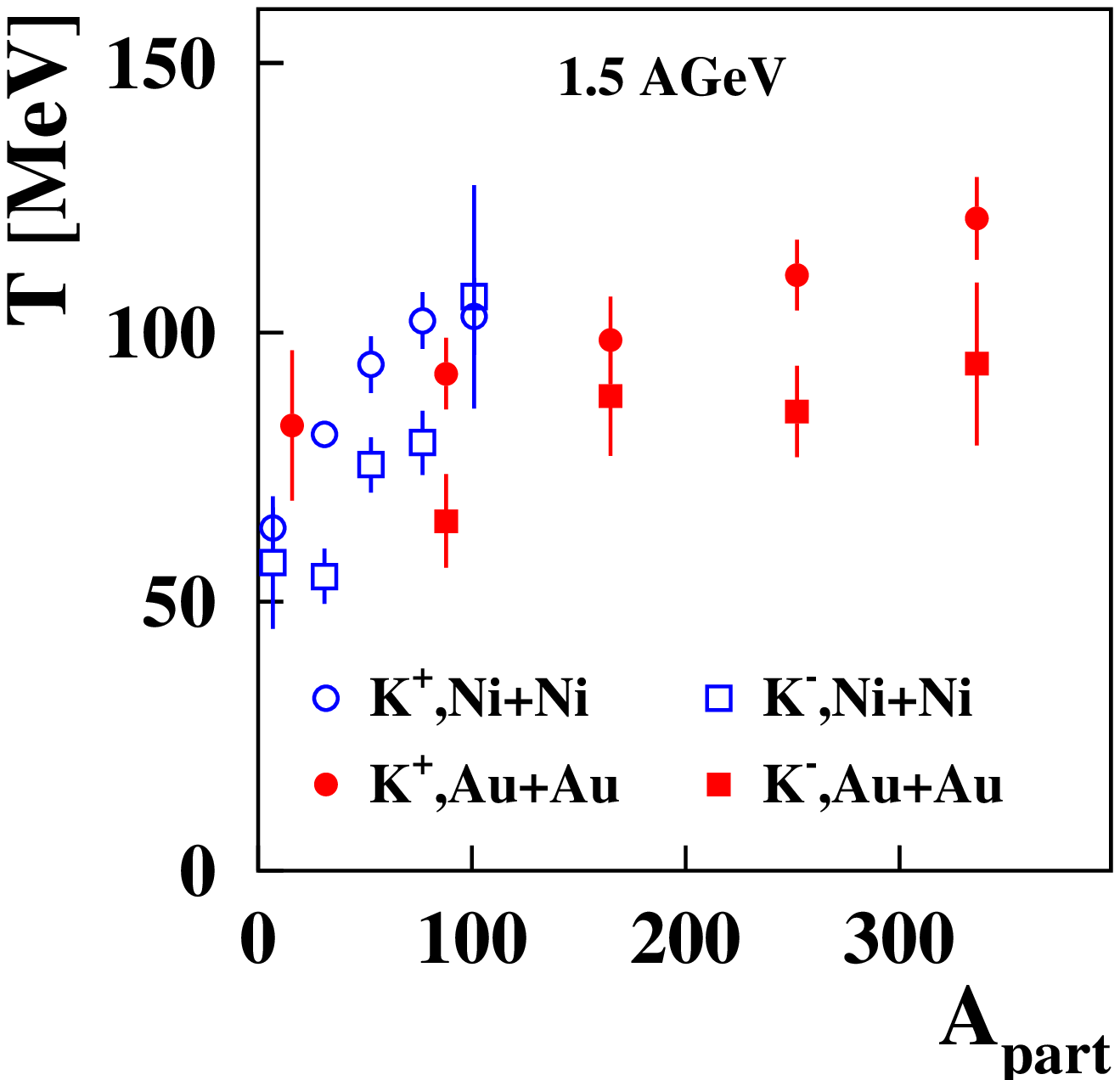,width=7cm}
\epsfig{file=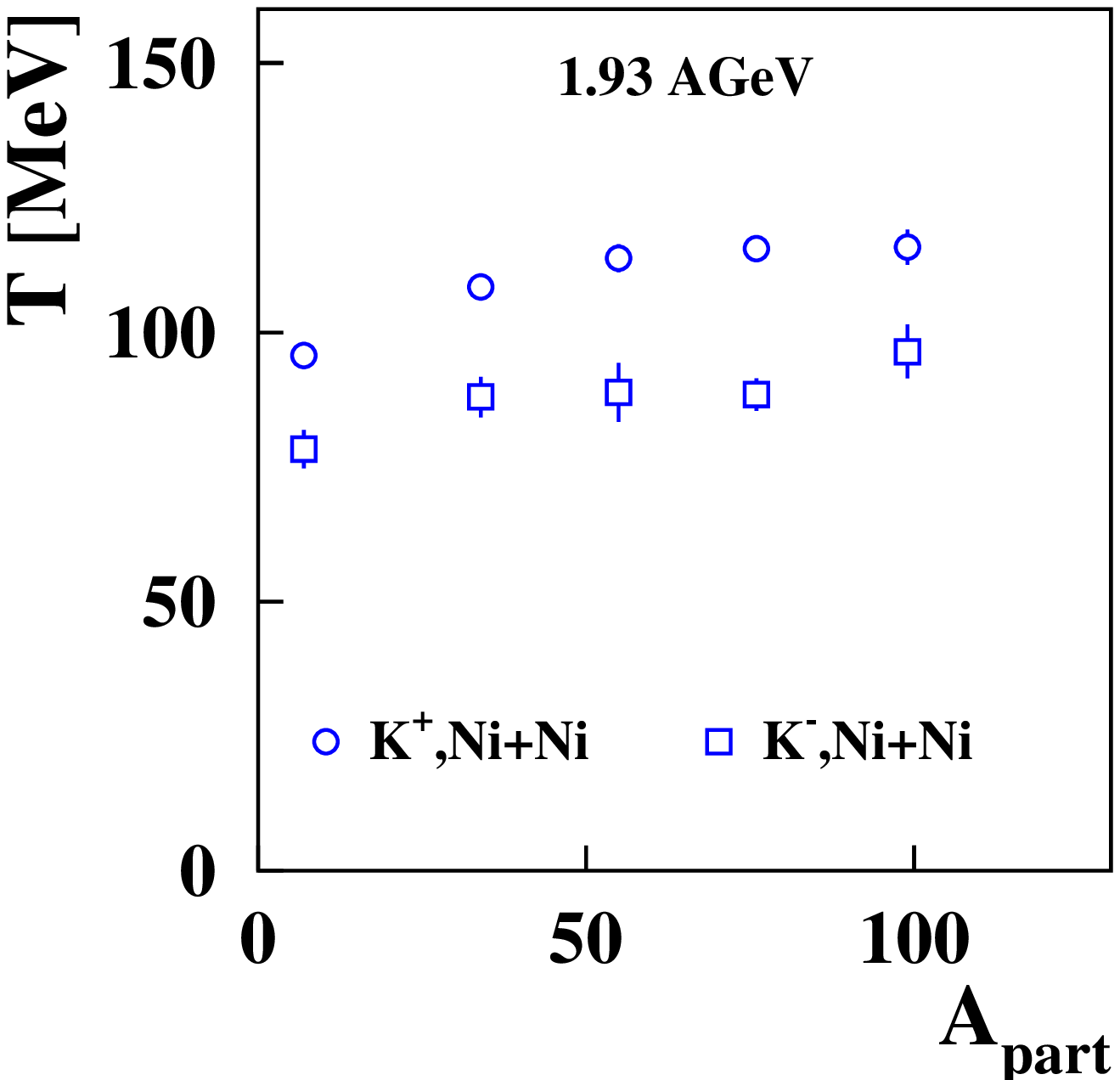,width=7cm}
 \caption{(Color online)
 Inverse slope parameters of the energy distributions of \kap
 and of \kam at $1.5$~\AGeV (upper part) and at $1.93$~\AGeV
 (lower part) as a function of \Apart[].}
\label{Au_Ni_slopes}
\end{figure}

The resulting inverse slope parameters $T$ for
\kap and \kam mesons are shown
in Fig.~\ref{Au_Ni_slopes} as a function of \Apart[].
The upper part displays \mbox{Au+Au} and \mbox{Ni+Ni}
collisions at 1.5~\AGeV[], the lower part
\mbox{Ni+Ni} collisions at 1.93~\AGeV[].
In Fig.~\ref{Au_Ni_slopes} two distinct features can be observed:

(i) The inverse slope parameters increase with centrality
for both particle species for both collision systems and both
beam energies.

(ii) The inverse slope parameters of the \kap spectra are larger
than those of the \kam spectra for both systems, at both energies
and for all centralities. This is discussed in detail in
Sect.~\ref{discussion_freezeout}.

\subsubsection{Polar Angle Distributions}

Another observable showing a distinct difference between \kap and
\kam is their polar angle emission pattern.
Deviations of the angular distributions from isotropy can be
studied by the ratio $\sigma_{\rm inv}(\theta_{\rm
c.m.})$/$\sigma_{\rm inv}(90^\circ)$ as a function of
$\cos(\theta_{\rm c.m.})$ as demonstrated in Sect.~III~A.
Here, $\sigma_{\rm inv}(\theta_{\rm c.m.})$ are the invariant
particle production cross sections measured at polar angles
$\theta_{\rm c.m.}$ in the center-of-momentum frame and
$\sigma_{\rm inv}(90^\circ)$ is deduced from Maxwell-Boltzmann fits
to the midrapidity spectra. These spectra have been obtained for the
centrality-dependent data in the same way as described for the
inclusive data in Sect.~\ref{results_inclusive}. For an isotropic
distribution this ratio would be constant and identical to 1.

Due to limited statistics we divided the data sets into two
centrality bins only: non-central collisions
($15 \, - \, 100~\%$ of $\sigma_{\rm r}$ for \mbox{Ni+Ni},
$18.1 \, - \, 100~\%$ of $\sigma_{\rm r}$ for \mbox{Au+Au}) and
near-central collisions
($0 \, - \, 15~\%$ of $\sigma_{\rm r}$ for \mbox{Ni+Ni},
$0 \, - \, 18.1~\%$ of $\sigma_{\rm r}$ for \mbox{Au+Au}).

\begin{figure}
 \epsfig{file=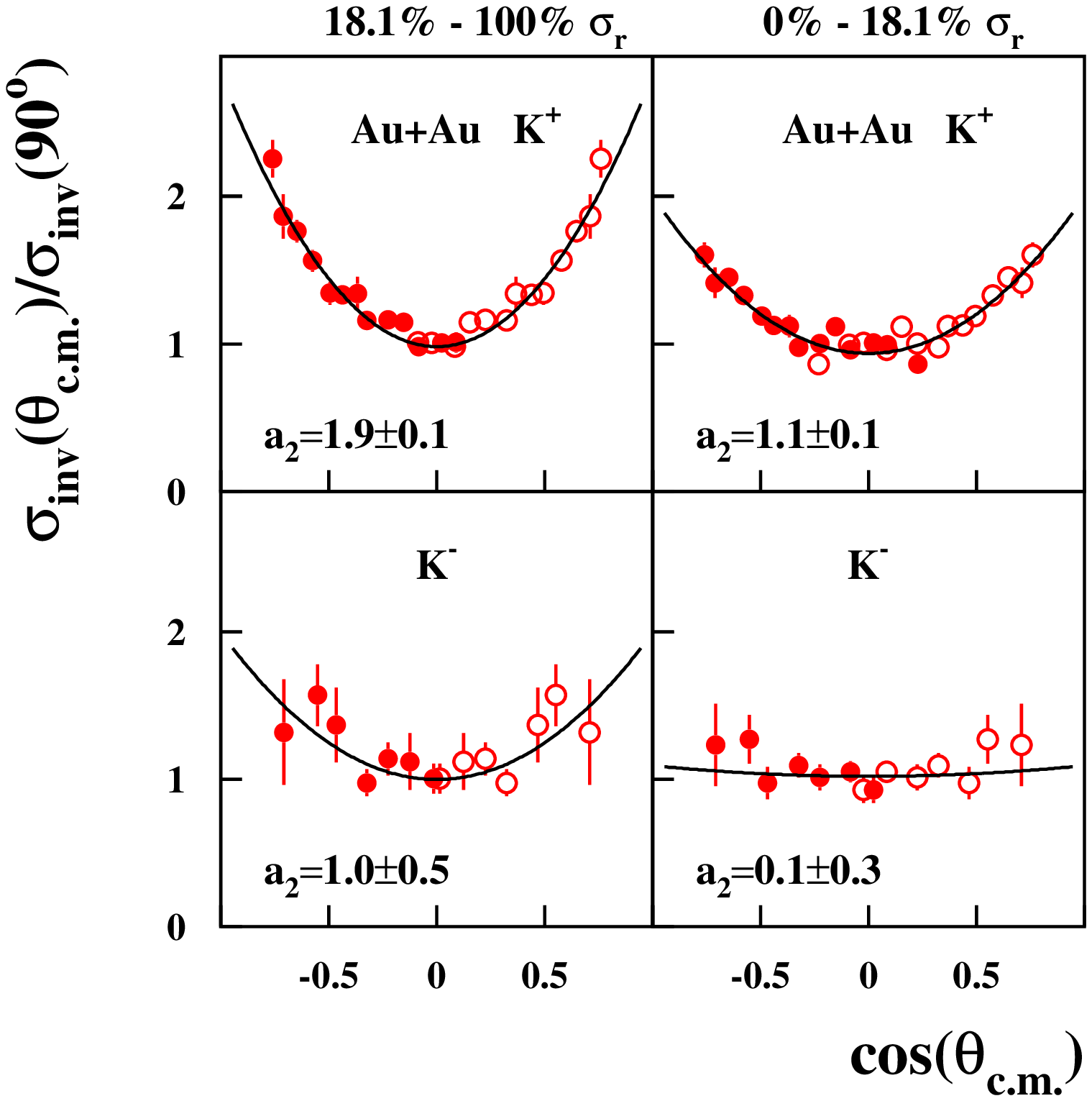,width=7.8cm}
 \epsfig{file=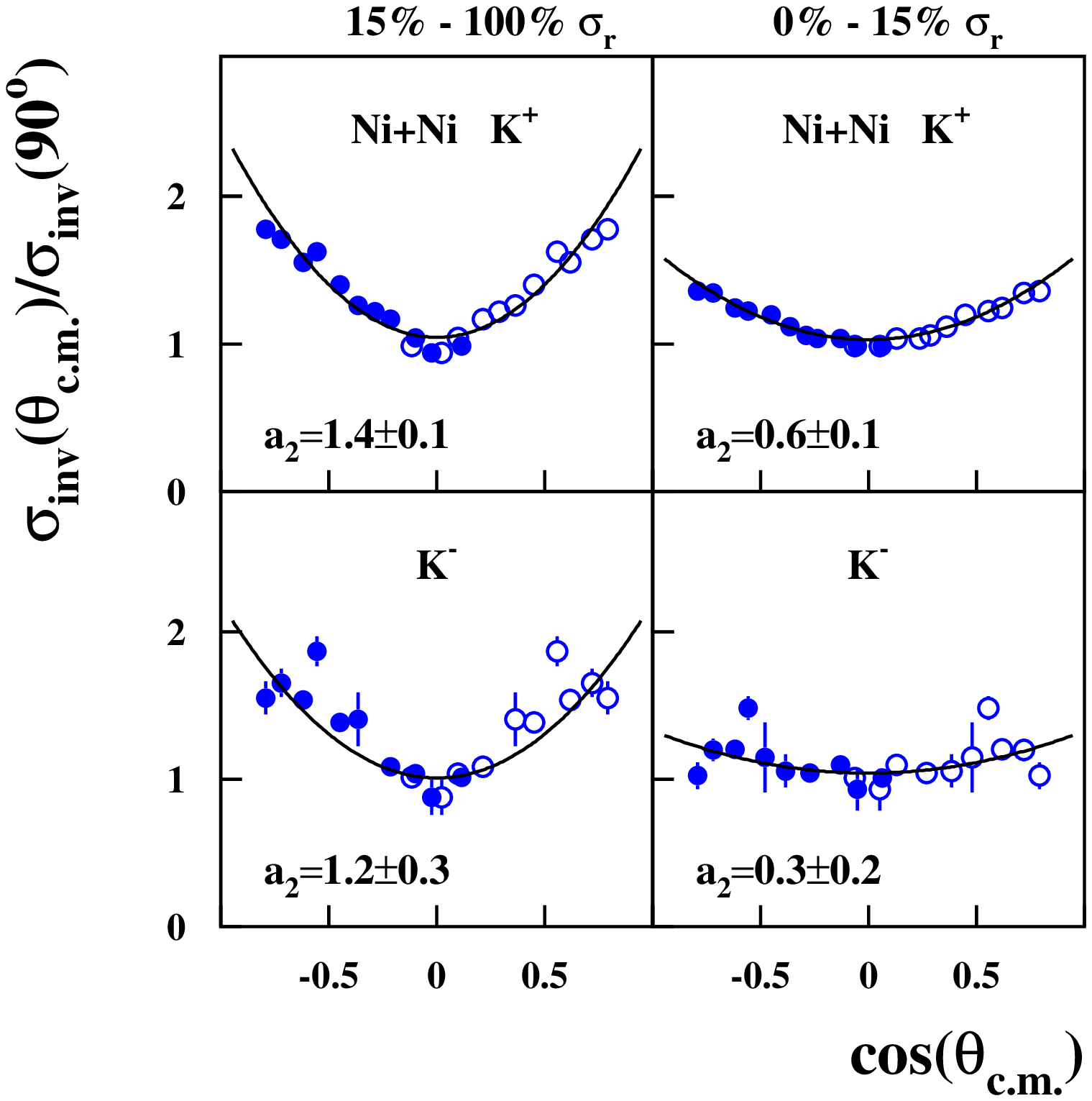,width=7.8cm}
 \caption{(Color online) 
  Polar angle distributions as measured in \mbox{Au+Au}
  at $1.5$~\AGeV
  (upper part) and in \mbox{Ni+Ni} at $1.93$~\AGeV (lower part).
  Full symbols
  denote measured data, open symbols are reflected at
  $\theta_{\rm{c.m.}} = 90^{\circ}$. The lines represent fits according
  to Eq.~(\ref{eq_3_2}).}
 \label{polar}
\end{figure}

Figure \ref{polar} displays the polar angle distributions for
\mbox{Au+Au} at $1.5$~\AGeV (upper part) and for \mbox{Ni+Ni} at
$1.93$~\AGeV (lower part). The upper panels of both parts of the
figure show \kap[], the lower panels \kam[], both for non-central
(left) and near-central collisions (right).

As for the inclusive data the distributions have been fitted with
a quadratic dependence on
$\cos(\theta_{\rm c.m.})$  according to Eq.~(\ref{eq_3_2}) to quantify
the anisotropy. The solid lines and the values for the parameter
$a_2$ in Fig.~\ref{polar} represent the results of these fits.
For both systems the \kam mesons exhibit a nearly isotropic emission
pattern in near central collisions whereas the emission of the
\kap mesons is forward-backward peaked.

\section{Discussion}
\label{discussion}

In this section we discuss the observed centrality
and system-size dependences of the \kap and of the
\kam multiplicities in the context of the production
mechanisms of these particles, we address the dynamics of the emission
of the two particle species in terms of
energy spectra and polar angle distributions,
and we extract information on the stiffness
of the nuclear equation of state by comparing the
\kap multiplicities from different collision
systems to recent transport model calculations.

\subsection{The Connection between the $\bf{K^{\boldsymbol{-}}}$
and the $\bf{K^{\boldsymbol{+}}}$ Production}
\label{discussion_coupling_kmkp}

As presented in Sect.~\ref{results_centrality}
the production yields of \kap and of \kam mesons exhibit a
very similar dependence on the collision centrality.
Figure~\ref{Au_ratio} shows that the multiplicities of both
kaon species exhibit the same rise with the number of
participating nucleons \Apart despite the fact that the
thresholds for the production of the two particles species
in binary NN-collisions are very different.
This is observed in \mbox{Au+Au} as well as in
\mbox{Ni+Ni} collisions.

\begin{figure}
\epsfig{file=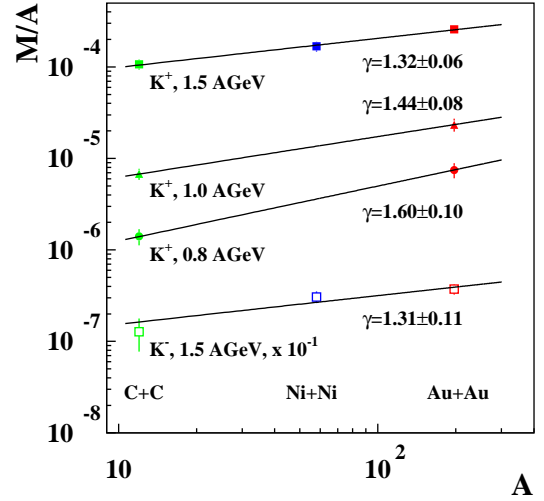,width=8cm}
 \caption{(Color online) 
  Multiplicities per mass number of the collision system
  $M/A$ as a function of $A$ for \mbox{C+C}, \mbox{Ni+Ni},
  and \mbox{Au+Au}. The lines represent the function
  $M \sim A^{\gamma}$ fitted to the data.}
\label{k_sigma_a}
\end{figure}

Figure~\ref{k_sigma_a} shows the multiplicities of \kap mesons
from inclusive
reactions as a function of the system size $A$ at several incident
energies as well as those of \kam mesons at $1.5$~\AGeV[].
To interpolate between measured data points in case of slight
differences in the effective beam energies due to different energy
losses in the respective targets the fits
to the excitation functions according to Eq.~(\ref{eq_kolomeitsev})
as shown in Fig.~\ref{k_sigma_Energy} have been used.
The lines in Fig.~\ref{k_sigma_a} are functions
$M \sim A^{\gamma}$ fitted to the data with the
resulting values for $\gamma$ given in the figure.
Please note: To distinguish the two approaches, we use
the exponent $\gamma$ to quantify the rise of the multiplicities
from inclusive reactions as a function of the system size $A$ and
the exponent $\alpha$ for the rise with the number of
participating nucleons \Apart as determined from the analysis of
the centrality dependence in Sect.~\ref{results_centrality}.

For \kap at $1.5$~\AGeV $\gamma = 1.32 \pm 0.06$ is extracted
which is almost identical to the value of $\alpha = 1.34 \pm 0.16$
as determined from the dependence of the multiplicity $M$ on
\Apart for the \kap production in \mbox{Au+Au} at the
same energy as shown in Fig.~\ref{Au_ratio}.
As can be seen in Fig.~\ref{k_sigma_a} $\gamma$ increases towards
lower incident energies. This is in good agreement with the
assumption of multiple collisions being needed to accumulate the
necessary energy for the \kap production at beam energies below
the threshold in binary nucleon-nucleon collisions. The lower the
incident energy the more collisions seem to be needed. Since the
densities reached in heavy reaction systems are significantly
larger than in light systems the difference between the \kap yield
in \mbox{C+C} and \mbox{Au+Au} increases with decreasing beam
energy. The same holds for the differences between peripheral and
central collisions in which different levels of baryon densities
are created. The similarity between the dependencies on \Apart and
on $A$ is therefore not astonishing. These observations will be
used in Sect.~\ref{discussion_eos} to extract the stiffness of the
nuclear equation of state.

The yield of the \kam mesons
as a function of the system size $A$
increases with $\gamma = 1.31 \pm 0.11$ at $1.5$~\AGeV. This is
roughly the same rise as for the dependence on \Apart in
\mbox{Au+Au} which was determined to be $\alpha = 1.22 \pm 0.27$
(see Sect.~\ref{results_centrality}). As in the case of the \Apart
dependence the rises of the \kam and of the \kap multiplicities
with $A$ are rather similar although the \kam production happens much
farther below its respective nucleon-nucleon threshold ($2.5$~GeV)
than the \kap production ($1.58$~GeV). In the case of the analysis
of the centrality dependence at $1.5$~\AGeV not only the same rise
of the  \kap and of the \kam multiplicities with centrality  was
observed but also a rough agreement between the data measured in
\mbox{Au+Au} and \mbox{Ni+Ni} (see Fig.~\ref{Au_ratio}).

\begin{figure}
\epsfig{file=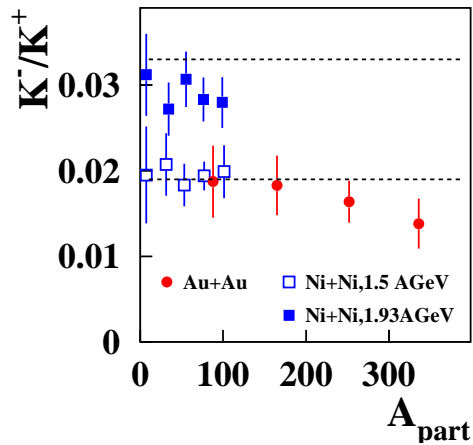,width=7.5cm}
 \caption{(Color online) \mbox{\kam[]/\kap} ratio as a function of \Apart
  for \mbox{Au+Au} at $1.5$~\AGeV (full circles)
  and for \mbox{Ni+Ni} at $1.5$~\AGeV (open squares)
  as well as at  $1.93$~\AGeV (full squares).
  The dashed lines denote the \mbox{\kam[]/\kap} ratios
  as calculated within the statistical model \cite{CLE99}.}
\label{kmkp_apart}
\end{figure}

The similar rise with \Apart yields a rather constant
\mbox{\kam[]/\kap} ratio as a function of \Apart as can be seen
in Fig.~\ref{kmkp_apart} for three cases.
At $1.5$~\AGeV the ratios for \mbox{Au+Au}
and for \mbox{Ni+Ni} are the same.
The dashed lines denote the \mbox{\kam[]/\kap} ratios
as calculated within the statistical model \cite{CLE99}.

The similar rise of both, \kap and \kam[], as a function of
the collision centrality as
shown in Fig.~\ref{Au_ratio} and as a function of the system size
(see Fig.~\ref{k_sigma_a}) suggests that the
production mechanisms of the two kaon species might be linked.
As already suggested in
\cite{ko84} and supported by transport model calculations
\cite{Hart03,cass_brat}, the \kam in heavy-ion collisions at SIS
energies are mainly produced via the strangeness-exchange reaction
$\pi {\rm Y} \rightleftharpoons {\rm K^-} {\rm N}$, with Y
denoting the hyperons ${\rm \Lambda}$ and ${\rm \Sigma}$.
On the other hand, strangeness has to be conserved when
producing these hyperons and the
energetically most favorable way is to produce them together with
\kap (and K$^0$) mesons (associate production). Thus the
production of \kap and of \kam mesons is coupled via the
strangeness-exchange reaction and the \kam inherit the same
dependence on the system size and on the collision centrality.
In \cite{Cl04} it was argued that the strangeness-exchange
channel reaches chemical equilibrium resulting
in the \mbox{\kam[]/\kap} ratio to be
proportional to the pion density and that
such a proportionality was observed for beam energies
lower than approximately $10$~\AGeV as reached at the SIS and at
the  AGS (Alternating Gradient Synchrotron) accelerators
at the GSI and at the Brookhaven National Laboratory (BNL).

\subsection{The Dynamics of the
$\bf{K^{\boldsymbol{+}}}$ and of the $\bf{K^{\boldsymbol{-}}}$ Emission}
\label{discussion_freezeout}

The strangeness exchange reaction couples the yields of the \kam
and of the \kap mesons as discussed in
Sect.~\ref{discussion_coupling_kmkp}.
On the other hand, \kap and \kam show rather distinct differences
in observables like energy spectra or polar angle distributions
which are sensitive to the dynamics of the particle emission.

\begin{figure}
\epsfig{file=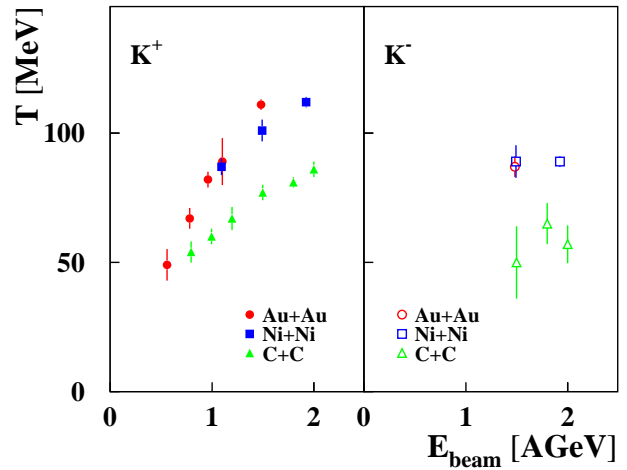,width=10cm}
 \caption{(Color online) Inverse slope parameters $T$ as determined from
 the energy spectra
 of \kap (left hand side) and of \kam (right hand side)
 for inclusive \mbox{C+C},
 \mbox{Ni+Ni}, and \mbox{Au+Au} collisions
 as a function of the beam energy.}
\label{slopes}
\end{figure}

Figure~\ref{slopes} presents the inverse slope parameters
$T_{\rm sf}$ (see Table~\ref{cross_table}) as determined
by the simultaneous fits  to the momentum distributions of the
production cross sections. They are shown as a function of the
incident energy for inclusive \mbox{C+C}, \mbox{Ni+Ni}
and \mbox{Au+Au} reactions, on the left hand side for \kap[],
on the right hand side for \kam[].
The inverse slope parameters  are higher for heavier systems and
in case of the \kap they
rise with increasing incident energies.
They are always higher for \kap than for \kam mesons
at the same beam energy.

\begin{figure}
\epsfig{file=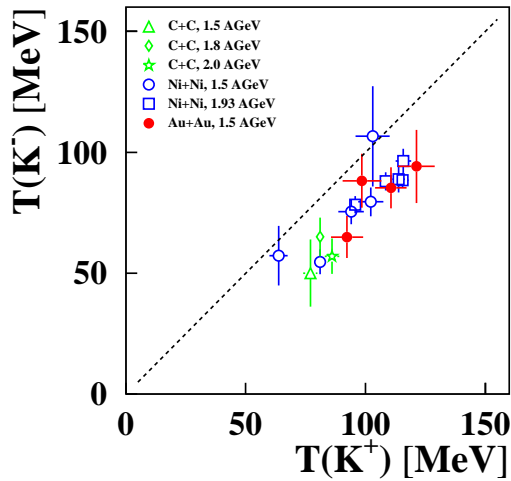,width=8cm}
 \caption{(Color online) Relation between the inverse slope parameters
  of the energy distributions
  of \kam and of \kap in various collision systems
  and at different incident energies (see text for details).}
\label{fig_sis_slope_kamkap}
\end{figure}

The same trend is as well observed as a function of the collision
centrality as shown in Fig.~\ref{Au_Ni_slopes} in
Sect.~\ref{results_centrality}. Figure~\ref{fig_sis_slope_kamkap}
shows the correlation between the inverse slope parameters $T({\rm
K^-})$ of the \kam mesons and those of the \kap mesons $T({\rm
K^+})$ measured in the same collision system and at the same
incident energy. For \mbox{C+C} the results from inclusive
collisions at $1.5$, $1.8$, and $2.0$~\AGeV are shown. For
\mbox{Ni+Ni} at $1.5$ and at $1.93$~\AGeV as well as for
\mbox{Au+Au} at $1.5$~\AGeV the results of the
centrality-dependent analyses are shown.
The dashed line indicates equal inverse slope parameters for \kam
and for \kap[]. The data clearly deviate from this line. The
inverse slope parameters of the \kap are about 15 - 25 MeV higher
than those of the \kam[], independent of the collision centrality,
of the collision system and of the beam energy (within the
measured energy range from $1.5$ to $2.0$~\AGeV[]).

While the measured \mbox{\kam[]/\kap}  ratio agrees well with
statistical-model calculations~\cite{CLE99}, the different values
for the inverse slope parameters for \kap and for \kam
clearly contradict the assumption of a simultaneous freeze out of both kaon
species. For \kam mesons the chemical and the kinetic freeze out coincide as
nearly no elastic scattering occurs due to the strong absorption.
If the kinetic decoupling of the \kap is at a higher ``temperature''
as the chemical freeze out of the \kam[], they cannot have a unique
chemical decoupling.

Different inverse slope parameters for \kap and for \kam mesons have been
proposed to result from the influence of the repulsive and of the
attractive KN potentials in early Boltzmann-Uehling-Uhlenbeck
(BUU) and in Relativistic-Quantum-Molecular-Dynamics (RQMD) transport-model
calculations~\cite{cass_brat,li_brown,wang}.
The experimentally observed difference between the
inverse slope parameters of \kap and of \kam mesons is
about the same for all reaction systems from \mbox{C+C} to
\mbox{Au+Au} collisions as well as for all collision centralities
and hence for very different densities
inside the collision zone. The KN potentials on the other hand are
predicted to have a strong dependence on the density~\cite{schaf}.
Yet, the data do not show such a dependence. This is a hint, that
other effects besides the KN potentials might as well be important
for the explanation of the different inverse slope parameters of
the \kap and of the \kam mesons.

Comparisons of data from the KaoS-Collaboration to various
transport-model calculations have been shown in several
publications. Some of the more recent comparisons can e.g. be
found in~\cite{Fuchs_Rev,hart_habil,Mishra,cas_tol,larionov} and
in the references therein. Another comprehensive report is in
preparation~\cite{PR_Nantes}. A comparison of the various transport models
can be found in~\cite{Trento}.
Here, we concentrate on the description of spectra and angular
distribution and their sensitivity to in-medium modifications of
kaons in dense nuclear matter. This comparison will be short as
further developments are expected from the theoretical side. Most
transport models use parameterizations for the KN-potentials that
result in density-dependent \kap and \kam selfenergies
(see~\cite{schaf} for examples). A more recent concept to describe
the in-medium properties via spectral functions of the \kam by
Lutz et al.~\cite{lutz} has not been implemented into transport
models so far. Another approach using coupled-channel G-matrix
calculations~\cite{laura} has been used for the K$^-$N interaction
in the Hadron String Dynamics Model (HSD)~\cite{cas_tol}.

In the following we compare energy distributions and polar angle
distributions to results of calculations obtained with the Isospin
Quantum Molecular Dynamics model (IQMD)~\cite{hart_habil,IQMD} and
with the HSD model~\cite{cas_tol}. These calculations have been
performed by the authors of the respective codes analyzing the
results within the experimental acceptance of the KaoS
measurements.

\begin{figure}[h]
\epsfig{file=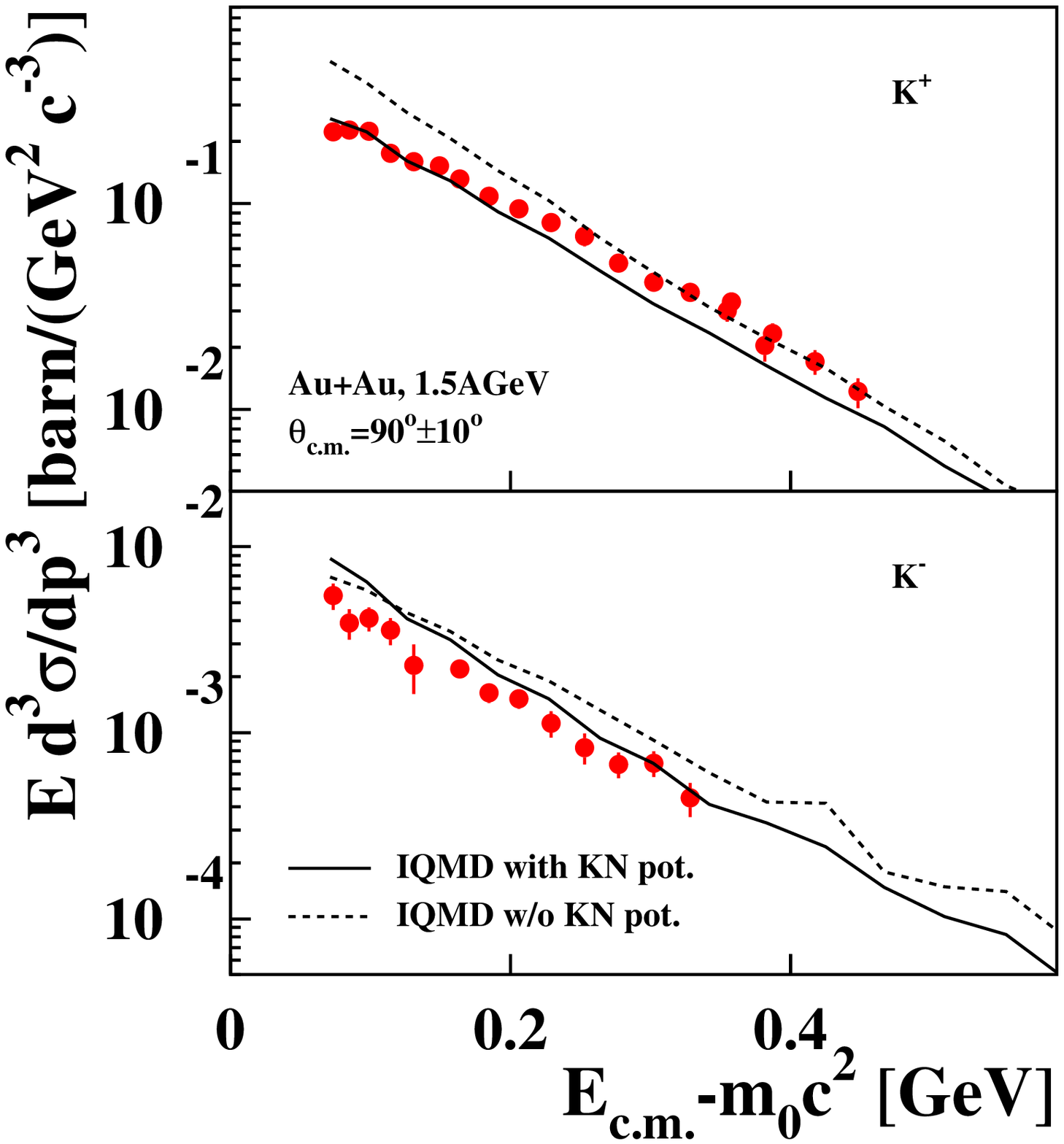,width=7.5cm}
\epsfig{file=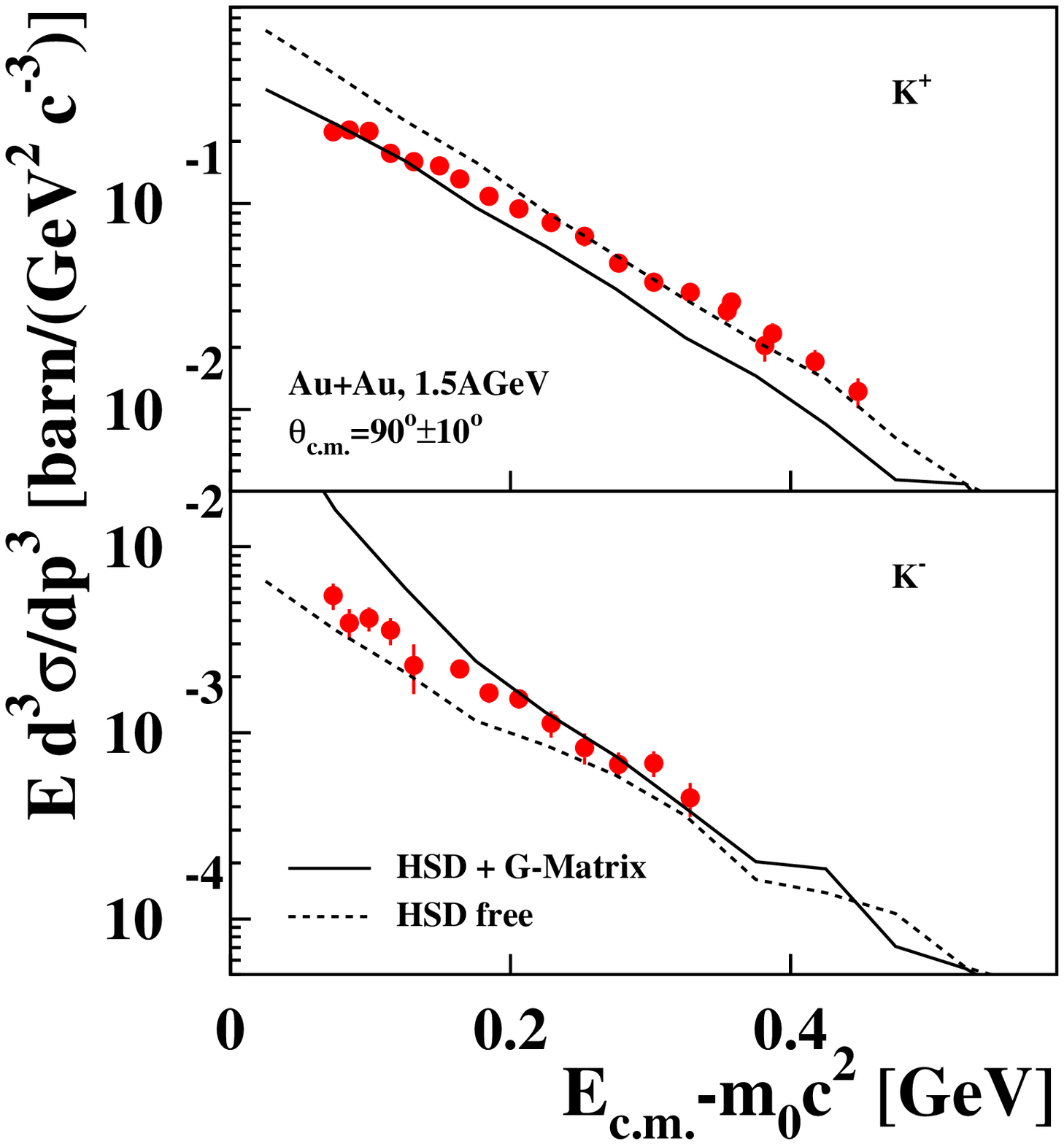,width=7.5cm}
 \caption{(Color online) Comparison of the energy distributions of the
 invariant cross sections of \kap and of \kam  in inclusive \mbox{Au+Au}
 at $1.5$~\AGeV
 with two different transport-model calculations, IQMD~\cite{hart_habil}
 (upper part) and HSD~\cite{cas_tol} (lower part).}
\label{auspec}
\end{figure}

Figure~\ref{auspec} displays the invariant cross sections
for \kap and for \kam mesons as a function of their kinetic energy
for Au+Au collisions at 1.5~\AGeV[].
In the upper part the comparison to IQMD calculations
is shown, in the lower part the comparison to
HSD results.
In both calculations the energy distributions of the \kap are
slightly steeper than for the measured data. The yields favor the
option with a repulsive K$^+$N potential. For the \kam[], the
slopes of both model calculations without an in-medium
K$^-$N interaction agree rather well with the data. The comparison
of the absolute yields, however, does not allow to draw a
conclusion  as the two calculations differ strongly. The
discrepancy is only seen for \kam[], while the results of the two
transport models agree rather well with each other for the \kap[].

As has been demonstrated, for the experimental data the inverse slope
parameters of the \kap are significantly larger than those of the
\kam[]. This trend is seen as well in the transport model
calculations, rather pronounced in those calculations including in-medium
KN interactions but already visible without them.

\begin{figure}[h]
\epsfig{file=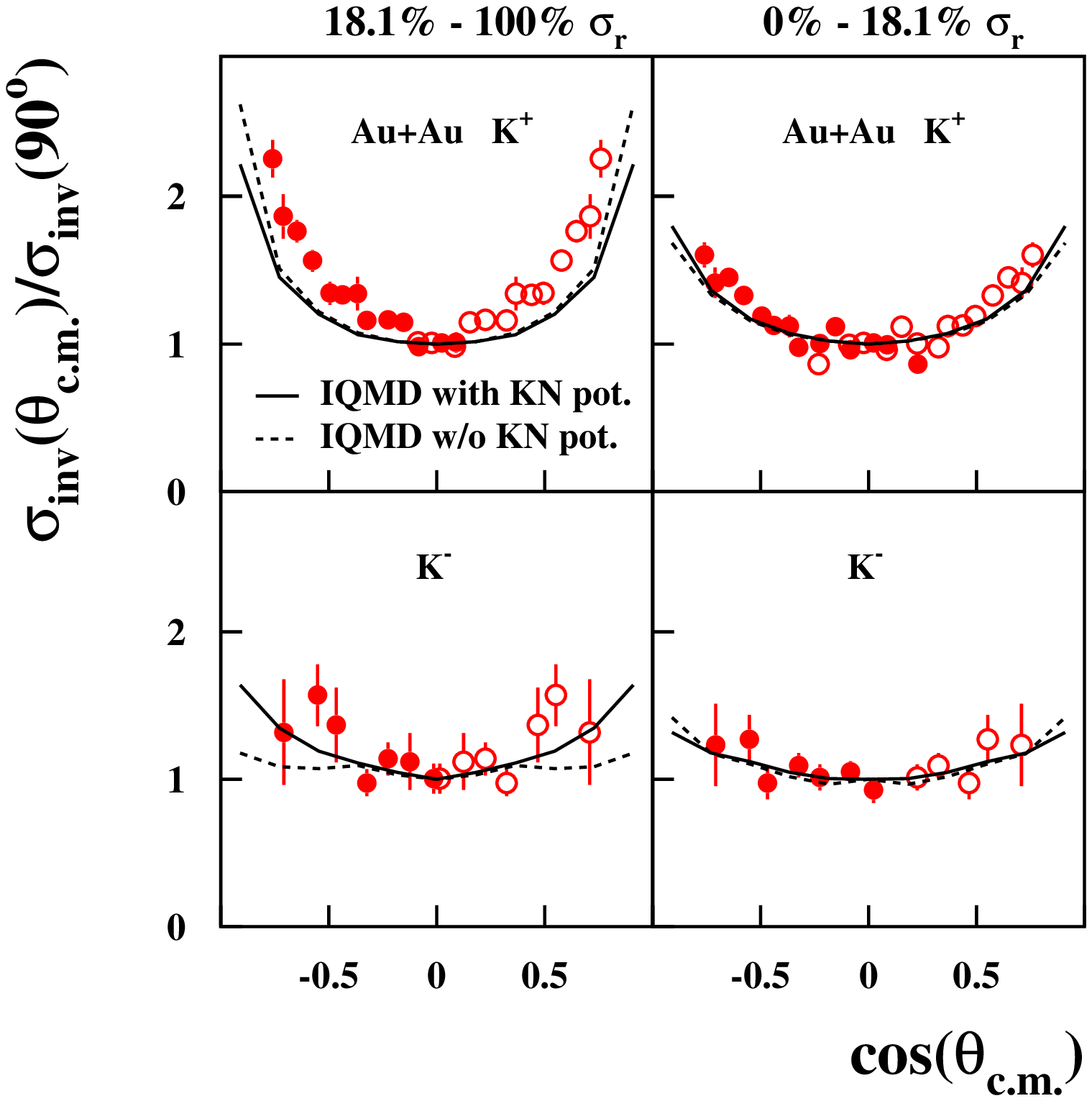,width=7.5cm}
\epsfig{file=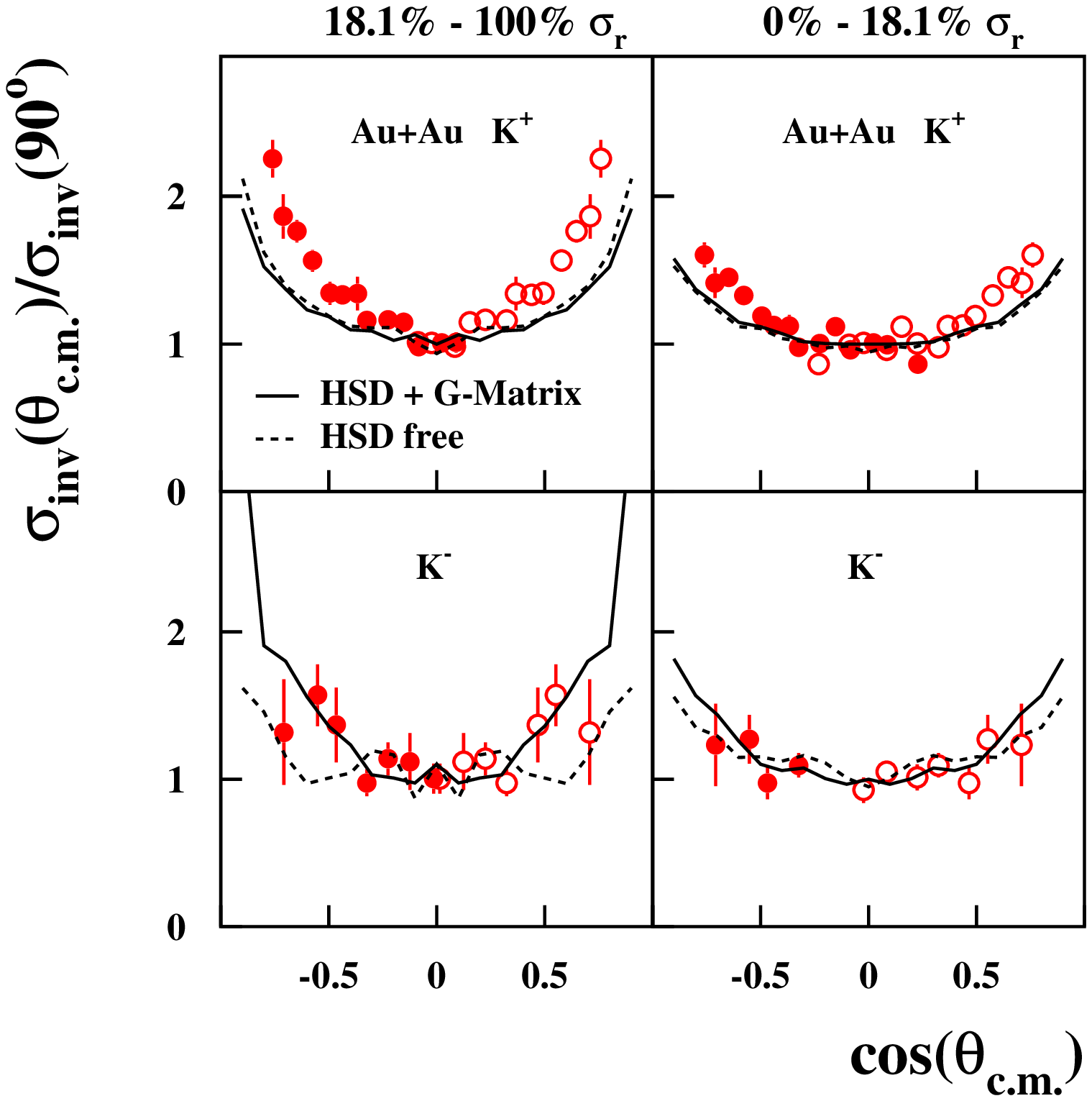,width=7.5cm}
 \caption{(Color online) Polar angle distributions of \kap and of \kam
  in \mbox{Au+Au} at $1.5$~\AGeV[]. The lines
  are the results of transport-model calculations,
  in the upper part using IQMD~\cite{hart_habil},
  in the lower part using HSD~\cite{cas_tol}.}
\label{angdis_data_calc}
\end{figure}

Figure~\ref{angdis_data_calc} compares the angular emission
patterns of \kap and of \kam in \mbox{Au+Au} at $1.5$~\AGeV[],
normalized to the yield at $\theta_{\rm c.m.} = 90^\circ$, to the
results of the transport models IQMD and HSD. For the \kam mesons
from near-central collisions ($0\%$ - $18.1\%~\sigma_{\rm r}$) a
rather flat distribution is observed, whereas the \kap are
preferentially emitted to forward and backward angles. The
measured data are rather well described by both models. However,
as can be seen in Fig.~\ref{angdis_data_calc} this observable is
hardly sensitive to the choice of the KN in-medium interaction.

\begin{figure}
\epsfig{file=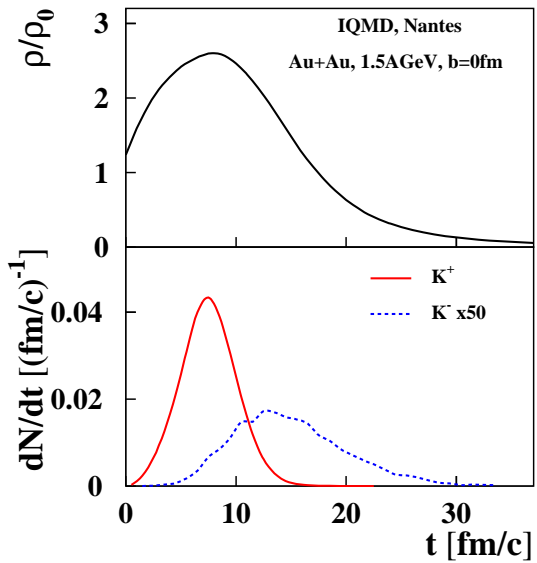,width=8cm}
 \caption{(Color online) 
  Results of IQMD calculations on the time evolution of central
  ($b=0$~fm) \mbox{Au+Au} collisions at $1.5$~\AGeV[].
  Upper panel: density normalized to normal nuclear matter density
  $\rho/\rho_0$ as a function of the time.
  Lower panel: rate of emitted \kap and \kam mesons  as a function
  of their production time.}
\label{IQMD_time}
\end{figure}

The differences in the inverse slope
parameters and in the polar angle distributions
of the \kap and of the \kam mesons
may as well be influenced by different emission
times of the two particle species. This is demonstrated in
Fig.~\ref{IQMD_time} showing the results of IQMD calculations.
The upper panel shows the density $\rho$ reached in the collision zone
normalized to normal nuclear matter density $\rho_0$ as
a function of time. The lower panel shows
the rate of emitted \kap and \kam mesons as a function
of their creation time. It can clearly be seen that
according to this transport-model calculation
those \kam leaving the reaction zone are created
at a rather late stage of the reaction, significantly
later than the \kap which are mainly created during
the high-density phase.
This difference in emission times is caused by the
strangeness-exchange reaction reabsorbing most of the
\kam produced during the high-density phase.
According to IQMD the primary production of both, \kap
and \kam mesons, is isotropic. As will be discussed in detail
in~\cite{PR_Nantes} the polar angle anisotropy of the \kap is mainly
caused by rescattering. The \kam on the contrary are emitted at a late stage
of the reaction when 
the spectator matter has moved away and
can not cause a significant anisotropy.

\subsection{The Nuclear Equation of State} \label{discussion_eos}

Positively charged kaons are expected to carry information on the
high density-phase of the heavy-ion collision due to two
reasons:

i) The \kap production occurs via multiple collisions e.g.~via
processes like ${\rm N N} \rightleftharpoons {\rm N} \Delta$
followed by a subsequent interaction of the $\Delta$ resonance
like ${\rm N} \Delta \rightleftharpoons {\rm K^+} \Lambda {\rm
N}$. These multi-step processes occur predominantly at higher
densities~\cite{Hart_NP94,Fuchs_Rev,hart_habil}.

ii) The rather large mean free path of the \kap ($\approx 5$~fm at
normal nuclear density $\rho_0$) reduces the probability of
further inelastic interactions prior to their emission. 
As a result of the
\kap production mechanism their yields are sensitive to 
the density reached
in the collision which is related to the stiffness of the nuclear
equation of state (EoS) as parameterized by the compression modulus
$K_{\rm N} $ defined as 
\begin{equation} \label{eq_comp_mod}
K_{\rm N} \, = \,
-V\frac{{\rm d}p}{{\rm d}V} \, = \,
9 \rho^2  \left. \frac{{\rm d}^2E/A(\rho,T)}{(\rm d\rho)^2}
\right|_{\rho=\rho_0} 
\end{equation}
which quantifies the curvature of $E/A(\rho,T)$ at
normal nuclear density $\rho_o$.

Figure~\ref{k_sigma_Energy} in Sect.~\ref{results_inclusive}
summarizes the multiplicities of \kap mesons as
determined in inclusive reactions of
\mbox{Au+Au}, \mbox{Ni+Ni}, and \mbox{C+C} and normalized to the
mass number $A$ of the respective colliding nuclei
as a function of the beam energy.
The \kap excitation functions for all three collision systems
rise strongly
as expected due to the proximity of the threshold in NN
collisions ($E_{\rm thr}=1.58$~GeV).
The multiplicities per $A$ are the higher the
heavier the collision system is. This reflects that the \kap mesons are
predominantly produced in multiple collisions which are more likely to
occur the higher the density is in the reaction.

Early transport-model calculations predicted that the \kap yield
in \mbox{Au+Au} collisions would be enhanced by a
factor of about 2 if a soft rather than a hard nuclear equation
of state is assumed \cite{aich,li_ko}. Recent calculations take
into account modifications of the kaon properties in the dense
nuclear medium leading to a repulsive ${\rm K^{+}N}$ potential
which depends on the baryonic density  \cite{schaf} and which
leads to a reduction of the calculated \kap yields.  To
disentangle these two competing effects we use the ratio of two
\kap excitation functions~\cite{sturm}, one from \mbox{C+C} and one from
\mbox{Au+Au}. The maximum baryonic density reached in
\mbox{Au+Au} reactions is about 2 -- 3 times normal nuclear matter
density while the increase in density in \mbox{C+C} collisions is
significantly less pronounced. Moreover, the maximum baryonic
density reached in \mbox{Au+Au} reactions depends  on the
compression modulus of nuclear matter $K_{\rm N}$
\cite{aichelin,li_ko} whereas in \mbox{C+C} collisions this
dependence is rather weak \cite{fuchs}. Hence, the ratio of the
\kap multiplicity per nucleon $M/A$ in \mbox{Au+Au} to the one in
\mbox{C+C} is expected to be sensitive to the compression modulus
$K_{\rm N}$. Furthermore, it provides the advantage that
systematic uncertainties within the experimental data are partly
cancelled. This ratio in addition has turned out to be hardly
affected by input quantities of the transport-model calculations
which are less well known, like cross sections of individual reaction
channels, the strength of the KN potentials, or the lifetime of
the $\Delta$ resonance, as has been systematically studied
in~\cite{Hart_eos}.

\begin{figure}
\vspace*{-1cm}
 \epsfig{file=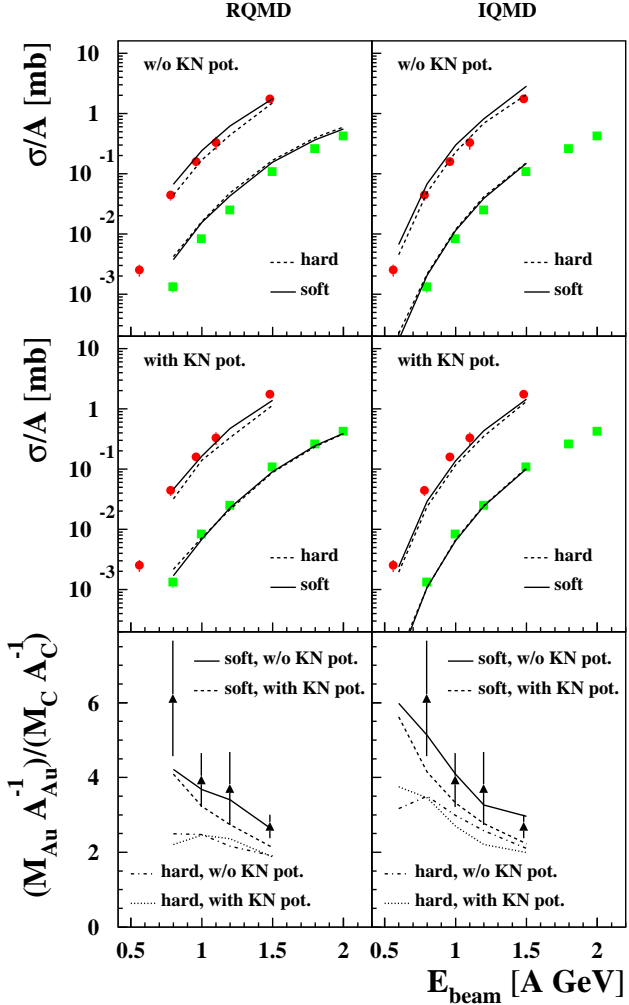,width=10cm}
  \caption{(Color online) Upper and middle panels:
  Comparison of the \kap excitation functions
  ($\sigma({\rm K^+})/A$) for \mbox{Au+Au}
  (red circles) and for \mbox{C+C} collisions (green squares)
  with RQMD~\cite{fuchs} (left hand side) and with
  IQMD calculations~\cite{Hart_eos} (right hand side).
  Solid lines depict a hard EoS, dashed lines
  depict a soft EoS.
  The upper panels show calculations without KN potentials,
  the middle panels calculations with KN potentials.
  Lower panels:
  Double ratio of the \kap multiplicities per mass number $M/A$
  in \mbox{Au+Au} divided by the one in \mbox{C+C} and
  the comparison to the various transport-model calculations.
  }
  \label{EOS}
\end{figure}

The upper and the middle panels of Fig.~\ref{EOS} show a detailed
comparison of the \kap excitation functions ($\sigma/A$ as a
function of $E_{\rm beam}$) for \mbox{Au+Au} 
and for \mbox{C+C} collisions 
with transport-model calculations. On the left hand
side of the figure the data are compared to results from 
RQMD calculations~\cite{fuchs}. 
On the right hand side a comparison to
results from IQMD~\cite{Hart_eos} calculations is
presented. The upper panels show calculations without KN
potentials, the middle panels those with KN potentials. Solid
lines denote calculations with a soft nuclear equation of state
($K_{\rm N} = 200$~MeV), dashed lines denote a hard EoS 
($K_{\rm N} = 380$~MeV).

The lower panels of Fig.~\ref{EOS} show the double ratio
$[M/A({\rm \mbox{Au+Au}})] / [M/A({\rm \mbox{C+C}})]$
as a function of the beam energy. Since due to the different energy
loss in the Au and in the C target the effective energies
for the \kap production are slightly different, the fits as displayed
in Fig.~\ref{k_sigma_Energy} were used for interpolation.
The error bars contain the statistical uncertainties
as well as those systematic errors that do not cancel by
calculating the double ratio (approximately 6\%), added quadratically.
The double ratios as determined
from the various transport-model calculations are shown as well.
Only the calculations using a soft EoS
are in agreement with the data.

%{\bf In order to estimate an error in determining the compression
%modulus $K_{\rm N}$, IQMD calculations have been performed
%studying the ratio $[M/A({\rm \mbox{Au+Au}})] / [M/A({\rm
%\mbox{C+C}})]$ at a given incident energy varying $K_{\rm N}$ as
%shown in Fig.~\ref{IQMD_ratio}. As expected the calculated ratios
%decrease with increasing stiffness. The experimental results are
%displayed as band. These figures allow to establish upper limits
%for $K_{\rm N}$, being $<$ 180 MeV (at 0.8 \AGeV[]) and $<$ 240
%MeV (at 1.0 \AGeV[]) assuming a KN potential. Higher limits are
%obtained without KN potential giving  $<$ 240 MeV (0.8 \AGeV[])
%and $<$ 315 MeV (1.0 \AGeV[]). These values do not contain
%uncertainties in the input of the calculations. For details see
%\cite{PR_Nantes}.}

\begin{figure}
 \epsfig{file=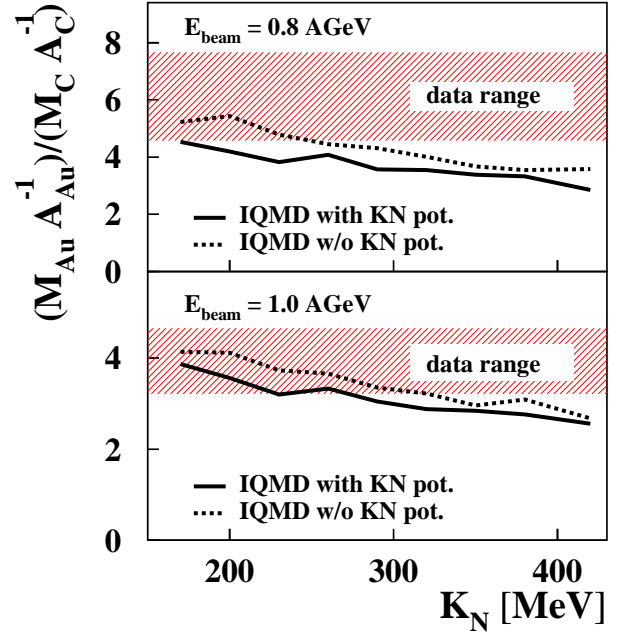,width=9cm}
  \caption{(Color online) The double ratio
  $[M/A({\rm \mbox{Au+Au}})] / [M/A({\rm \mbox{C+C}})]$
calculated within the IQMD model (with and without KN potential)
as a function of $K_{\rm N}$ at 0.8 and at 1.0 \AGeV[]. The
experimental values are given as bands and allow to estimate upper
limits for $K_{\rm N}$ as described in the text.}
  \label{IQMD_ratio}
\end{figure}

A reliable error estimate for the compression modulus
$K_{\rm N}$ strongly depends on the transport model calculations 
and their input. A detailed study of this topic will be subject of 
a theory publication \cite{PR_Nantes}. 
The sensitivity of the double ratio 
$[M/A({\rm \mbox{Au+Au}})] / [M/A({\rm \mbox{C+C}})]$  
to the stiffness of the EoS within the standard version of IQMD
is shown in Fig.~\ref{IQMD_ratio}. It compares the measured double
ratio at 0.8 and 1.0~\AGeV (shown as red bands) to results of 
IQMD calculations as a function of $K_{\rm N}$,
both with and without KN potentials. 
The figure allows to establish upper limits
for $K_{\rm N}$, being $<$ 180 MeV (at 0.8 \AGeV[]) and $<$ 240
MeV (at 1.0 \AGeV[]) assuming a KN potential. Higher limits are
obtained without KN potential resulting in $<$ 240 MeV (at 0.8 \AGeV[])
and $<$ 315 MeV (at 1.0 \AGeV[]). 

In Ref.~\cite{schmah} a detailed comparison of the \kap
production in the mass-asymmetric collision system \mbox{C+Au} as
well as in \mbox{Ni+Ni}, both measured at the Kaon Spectrometer,
with RQMD calculations is presented. Although the mean number of
nucleons participating in the reaction is rather similar in both
cases, the densities reached are significantly different. The
compression in \mbox{C+Au} hardly exceeds the values obtained in
\mbox{C+C} and therefore the \kap yield in the calculations does
not show a dependence on the stiffness of the EoS. For
\mbox{Ni+Ni} the compression is significantly higher and 
the difference in the calculated yields is about
$25\%$. Again, the RQMD calculation for $K_{\rm N}=200$~MeV is in good
agreement with the data.

\begin{figure}[h]
  \epsfig{file=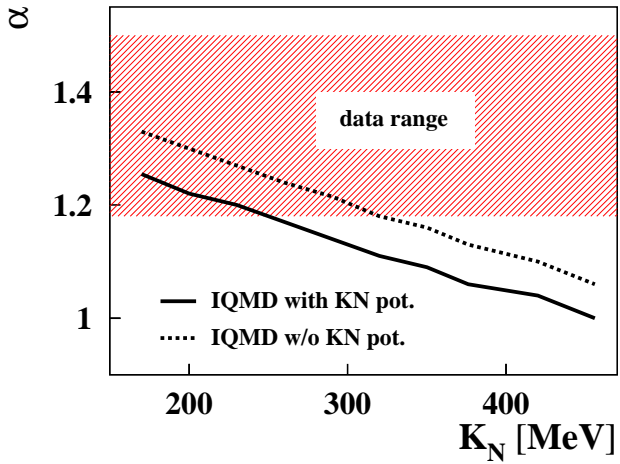,width=9.5cm}
  \caption{(Color online) Comparison of the exponent $\alpha$ from the fit of
    $M \sim A_{\rm part}^{\alpha}$ to the
    \kap multiplicities in \mbox{Au+Au} at $1.5$~\AGeV
    (shaded area)
    with IQMD calculations as a function of the compression
    modulus $K_{\rm N}$~\cite{Hart_eos}.}
  \label{alpha_iqmd}
\end{figure}

In Ref.~\cite{Hart_eos} the centrality dependence of the 
\kap production has been suggested as a further observable 
to extract the stiffness of the nuclear EoS 
from heavy-ion collisions. 
As shown in Fig.~\ref{Au_ratio} the \kap multiplicity $M$ in
\mbox{Au+Au} at $1.5$~\AGeV rises stronger than linear with the
number of participating nucleons \Apart[]. This is due to the
production via multiple collisions which is needed to accumulate the
necessary energy and due to the dependence of the number 
of collisions on the
density reached in the reaction zone. In Sect.~\ref{results_centrality} 
this rise has been quantified by a fit to the data of the form 
$M \sim A_{\rm part}^{\alpha}$ with $\alpha = 1.34 \pm 0.16$.
Figure~\ref{alpha_iqmd} compares this value (shaded area) to results
of IQMD transport-model calculations on the dependence of $\alpha$
on the compression modulus $K_{\rm N}$~\cite{Hart_eos,PR_Nantes}. 
Only values for $K_{\rm N}$ less than  $250$~MeV  are
compatible with the measured data assuming a KN potential, and
$K_{\rm N} \leq 320$ MeV for the case without a KN potential.

We would like to stress that two independent observables, the
centrality dependence of the \kap yields, shown in
Figs.~\ref{Au_ratio} and \ref{alpha_iqmd}, as well as the system-size
dependence presented in Figs.~\ref{k_sigma_a} and \ref{EOS}, 
yield very similar
results on the compressibility of nuclear matter when compared
to transport-model calculations. 
Both observables support a soft nuclear EoS
within the density regime explored by heavy-ion reactions
at beam energies between $0.6$ and $2.0$~\AGeV[].

\section{Summary}
\label{summary}

In this paper we have presented a comprehensive review of the
production of charged kaons in heavy-ion collisions at incident
energies ($0.6$ to $2.0$~\AGeV[]) below and at the respective
thresholds in NN collisions as measured with the Kaon Spectrometer
KaoS at GSI. This subject has been systematically studied  by
analyzing total production cross sections, energy distributions
and polar angle distributions as a function of the size of the
collision system, of the incident energy, and of the collision
centrality. The key observations and trends can be summarized as
follows:

\begin{itemize}

\item The multiplicities of both, \kap and \kam mesons,
per mass number $A$ of the collision system are
higher in heavy collision systems than in light systems.
This difference increases with decreasing beam energy.

\item The multiplicities per
number of participating nucleons \Apart 
of \kap and of \kam mesons  within the same
collision system rise
stronger than linearly with \Apart whereas the pion
multiplicity is proportional to \Apart[].
Moreover the rise is rather similar for \kap and for
\kam although the respective NN-thresholds for their
production are significantly different.

\item The \mbox{\kam[]/\kap} ratio is
almost constant as a function of the collision centrality.
At $1.5$~\AGeV this ratio is the same for \mbox{Au+Au}
and for \mbox{Ni+Ni} collisions.

\item The inverse slope parameters of the energy distributions
of \kap and of \kam mesons are higher in heavy than
in light collision systems.

\item The inverse slope parameters of the energy distributions
of \kap mesons are about
$15$ to $25$~MeV higher than those of the \kam distributions.
This is observed for all
collision systems and for all centralities.

\item The polar angle distributions exhibit a forward-backward
rise which is more pronounced for \kap than for \kam mesons. \kam
mesons produced in central collisions are emitted almost isotropically.

\end{itemize}

From the systematics of these experimental results
and from detailed comparisons to transport model calculations
the following conclusions on the properties of
dense nuclear matter as created in heavy-ion collisions
and on the production mechanisms of \kap and of \kam mesons
can be drawn:

\begin{itemize}

\item {\it The \kam and the \kap yields are coupled by strangeness
exchange:} Despite their significantly different thresholds in
binary NN collisions the multiplicities of \kap and of \kam mesons
show the same dependence on the collision centrality. They are even
similar for different collision systems. This can be explained by
the \kam being predominantly produced via strangeness exchange
from hyperons which on the other hand are created together with
the \kap mesons. Strangeness exchange is predicted to be the main
contribution to the \kam production in heavy-ion collisions at SIS
energies by transport model calculations as well.

\item {\it \kap and \kam mesons exhibit different freeze-out
conditions:} Transport-model calculations predict different
emission times for \kap and for \kam mesons as a consequence of
the strangeness-exchange reaction. The \kam are continuously
produced and re-absorbed and finally leave the reaction zone much
later than the \kap mesons. This and the kinematics of the
strangeness-exchange process manifests itself in an isotropic
emission of the \kam in central collisions and by systematically
lower inverse slope parameters of the \kam energy distributions
compared to \kap.

\item {\it The nuclear equation of state is soft:} The increase of
$M({\rm K^+})/A$ with the size of the collision system $A$ points
towards a dependence of the \kap production on the density reached
in the collision. The  ratio of the \kap multiplicities in
\mbox{Au+Au} and \mbox{C+C} as a function of the incident energy
allows  to extract the compression modulus $K_{\rm N}$ of nuclear
matter by comparing to transport-model calculations. Only
calculations using a soft nuclear equation-of-state ($K_{\rm N}
\approx 200$~MeV) can describe the data. This conclusion is rather
insensitive to the various input parameters of such calculations.
A soft nuclear equation of state is further supported by comparing
the centrality dependence of the \kap multiplicities in
\mbox{Au+Au} collisions to transport model calculations.

\end{itemize}

Our results demonstrate the importance of the strangeness-exchange
reaction for the production and propagation of negatively charged
kaons in heavy-ion collisions at incident energies from $0.6$ to
$2$~\AGeV[], on the one hand coupling their yield to the
\kap-production, on the other hand causing a rather late emission
of the \kam[]. The production of positively charged kaons itself is
strongly linked to the high-density phase of a heavy-ion collision
allowing to determine that the equation-of-state of nuclear matter
is soft within the density regime explored by heavy-ion collisions
between $0.6$ and $2.0$~\AGeV[].

\hspace{1cm}

\noindent{\bf Acknowledgements}\\

We would like to thank all our colleagues with whom we had
numerous fruitful discussions over the past years. We explicitly
acknowledge the intense collaboration with the various theory
groups providing the results of the transport model calculations,
especially J. Aichelin, E. Bratkovskaya, W. Cassing, C. Fuchs, and
C. Hartnack. This work was supported by the German Federal
Government (BMBF), by the Polish Committee of Scientific Research
(No. 2P3B11515) and by the GSI fund for Universities.

\end{document}